\documentclass[sigconf]{acmart} %

\usepackage[inline, shortlabels]{enumitem}
\usepackage{color, colortbl}

\usepackage{xurl}

\usepackage{adjustbox}
\usepackage{multirow}
\usepackage[para,flushleft]{threeparttable}

\usepackage{filecontents}

\usepackage{algorithm} 
\usepackage{algpseudocode}

\usepackage{amsmath}

\usepackage{wasysym}
\usepackage{pifont}

\usepackage{multirow}
\usepackage{tabularx}

\usepackage[leftmargin=0.5em,vskip=0.4em]{quoting}

\usepackage{cancel}
\usepackage{ulem} %
\usepackage{tikz}

\usepackage{scalerel,graphicx,xparse}

\usepackage{hyperref}    %
\usepackage{cleveref}    %
\usepackage{soul}
\usepackage{booktabs}
\usepackage{diagbox}

\usepackage{enumitem}
\usepackage{listings}

\usepackage{subcaption}
\usepackage{bm}
\usepackage{multirow}
\usepackage{booktabs}
\usepackage{numprint}
\usepackage{tikz}

\newcommand*{\halfDiamond}{%
  \begin{tikzpicture}[baseline=0.2ex]
    \clip (0,0) -- (0.9ex,0.9ex) -- (0,1.8ex) -- (-0.9ex,0.9ex) -- cycle;
    \fill[black] (-1ex,0) rectangle (0,2ex);
    \draw (0,0) -- (0.9ex,0.9ex) -- (0,1.8ex) -- (-0.9ex,0.9ex) -- cycle;
  \end{tikzpicture}%
}

\newcommand*{\fullDiamond}{%
  \tikz\draw[black,fill=black] (0,0) -- (0.9ex,0.9ex) -- (0,1.8ex) -- (-0.9ex,0.9ex) -- cycle;
}

\newcommand{\empirical}[1]{\setlength{\fboxsep}{1pt}\fbox{#1}}
\renewcommand{\empirical}[1]{#1}


\usepackage{contour}
\usetikzlibrary{tikzmark}

\usepackage{balance}

\crefname{figure}{fig.}{fig.}
\Crefname{figure}{Fig.}{Fig.}
\crefformat{section}{\S#2#1#3}
\Crefformat{section}{\S#2#1#3}
\crefformat{subsection}{\S#2#1#3}
\Crefformat{subsection}{\S#2#1#3}
\crefformat{subsubsection}{\S#2#1#3}
\Crefformat{subsubsection}{\S#2#1#3}
\crefname{table}{tab.}{tab.}
\Crefname{table}{Tab.}{Tab.}

\newcommand{\pseudocomment}[1]{\textcolor[HTML]{6A9955}{\# #1}}

\AtBeginDocument{%
  }

\copyrightyear{2026}
\acmYear{2026}
\setcopyright{cc}
\setcctype{by}
\acmConference[CCS '26]{Proceedings of the 2026 ACM SIGSAC Conference on Computer and Communications Security}{November 15--19, 2026}{The Hague, Netherlands.}
\acmBooktitle{Proceedings of the 2026 ACM SIGSAC Conference on Computer and Communications Security (CCS '26), November 15--19, 2026, The Hague, Netherlands}
\acmISBN{979-8-4007-2871-6/2026/11}
\acmDOI{10.1145/3830454.3846573}
\begin{document}

\title[Omniscience for the Masses]{Omniscience for the Masses: New Threats in the Metaverse's Democratized World Creation} %

\author{Andrea Mengascini}
\affiliation{%
  \institution{CISPA Helmholtz Center for Information Security}
  \city{Saarbr\"ucken}
  \country{Germany}}
\email{andrea.mengascini@cispa.de}

\author{Ryan Aurelio}
\affiliation{%
  \institution{CISPA Helmholtz Center for Information Security}
  \city{Saarbr\"ucken}
  \country{Germany}}
\email{ryan.aurelio@cispa.de} %

\author{Jason Polakis}
\affiliation{%
  \institution{University of Illinois Chicago}
  \city{Chicago}
  \state{Illinois}
  \country{USA}}
\email{polakis@uic.edu}

\author{Giancarlo Pellegrino}
\affiliation{%
  \institution{CISPA Helmholtz Center for Information Security}
  \city{Saarbr\"ucken}
  \country{Germany}}
\email{pellegrino@cispa.de} %

\renewcommand{\shortauthors}{Mengascini et al.}

\begin{abstract} %
Metaverse platforms increasingly derive their success from user-generated \emph{virtual worlds}: self-contained social and interactive environments, which can be created by \emph{any} ordinary user and scale to billions of visits. Platforms such as Roblox, Horizon Worlds, and VRChat now host millions of creator-built worlds that govern how users see, hear, and interact with one another. While this model enables rapid growth and creativity, it fundamentally delegates control over social interactions and world behavior to untrusted users. In this paper, we present the first systematic security and privacy assessment of metaverse world creators. We survey 25 platforms that support user-created worlds and analyze their world-creation capabilities. Guided by this analysis, we design and implement five novel attacks that exploit creator-provided tools to violate spatial, visual, and auditory constraints in immersive environments, enabling covert user surveillance and manipulation without software vulnerabilities or developer-level privileges. We further show that five previously-proposed attacks can be replicated using only standard world-creation features. Finally, we find that existing platform vetting, runtime protections, and creator policies are insufficient to mitigate malicious world-creator behavior, revealing a fundamental mismatch between users' privacy expectations and the powers granted to world creators.
\end{abstract}



\maketitle

\section{Introduction}
\label{sec:introduction}

Metaverses are virtual online environments that allow users to interact, play
games, and communicate in an immersive manner, attracting hundreds of millions
of daily users~\cite{RDC2025R40:online,MetricsI65:online}. A key
driver behind their popularity is the growing support for user-generated content,
which has recently further evolved to allow users to create and share entirely
new \emph{virtual worlds}. Nowadays, popular platforms such as Roblox, Horizon
Worlds, and VRChat derive much of their value from millions of creator-built
worlds~\cite{MetricsI65:online}, where
individual worlds  %
can accumulate tens of billions of visits~\cite{MisteroO57:online}.

User-created virtual worlds are self-contained experiences such as social
gatherings, business meetings, mini-games, or event spaces that users can
join within a metaverse platform. Much like Discord channels or IRC rooms,
these customizable spaces are executed within the platform's infrastructure,
inheriting the platform's runtime and interaction model so creators can build
upon them without the need to start from scratch. To define a world's behavior
and game logic, creators leverage a range of tools offered by the platforms, from
standalone game development environments to in-app WYSIWYG editors supporting
visual programming. This shift toward metaverses with world-creation capabilities
has introduced a new role in the ecosystem that has not yet been studied: the
\emph{world creator}, whose design decisions directly affect the experience of
a vast number of users.

The security and privacy of metaverses have been studied extensively in recent
years, typically under a threat model that distinguishes between developers and
users. Malicious developers, with their privileged access, can exploit VR device
sensors to identify, profile, and track
users~\cite{Munilla_Garrido_Nair_Song_2024,Nair_Guo_Mattern_Wang_OBrien_Rosenberg_Song_2023}
or 
modify scene-orientation to hijack physical movements and cause
harm~\cite{Casey_Baggili_Yarramreddy_2021}. Malicious users, by contrast, are
assumed to have limited influence over the environment and instead rely on
client-side vulnerabilities, memory inspection, or network traffic analysis to
track~\cite{Mengascini_Aurelio_Pellegrino_2024}, surveil~\cite{Mengascini_Aurelio_Pellegrino_2024},
and monitor others~\cite{Su_Cai_Beeler_Dresel_Garcia_Grishchenko_Tian_Kruegel_Vigna}.
World creators do not fit either category. While they remain ordinary,
unverified users from the platform's perspective, they are entrusted with
fine-grained control over the runtime logic and interaction mechanisms of shared
social environments. This hybrid role places world creators in a blind spot of
existing threat models: they lack developer privileges, yet wield far greater
influence over others' experiences compared to traditional users, introducing
significant risks. %

These security and privacy risks are amplified by the very design choice that
makes metaverses feel immersive: platforms emulate real-world physical
constraints such as users' sight being blocked by virtual walls and doors, and
spatial audio decaying with distance. These cues shape not only how users move
and interact (e.g.,~one cannot walk through a wall), but also how they reason
about privacy within virtual spaces (e.g.,~conversations held behind closed
doors are presumed to be private)~\cite{limbago2025don,williamson2021proxemics}.
Prior work shows that when such cues are respected, users disclose freely; when
they are covertly violated, users feel betrayed and
exposed~\cite{sykownik2022something}. World creators operate precisely at this
boundary. The tools that allow them to construct secluded rooms or muted
zones can be creatively misused to \emph{stealthily} access, observe, and manipulate
avatar state and communication streams within those spaces. The 
mismatch between user expectation and capability---\emph{the assumption of
physical-world-like privacy versus the actual powers of creators}---renders the
security and privacy implications of the world-creator role both urgent and
critical.

In this paper, we present the first systematic investigation of the security and
privacy threats posed by metaverse world creators. As world-creation features
are relatively recent and unevenly documented, we begin with a systematic survey
of \empirical{25} metaverse platforms, including Horizon Worlds, Roblox, and
VRChat, that support user-created worlds, and categorize their world-creation
capabilities based on the expressiveness of their creation tools. Motivated by
our findings and the features provided by the platforms, we conduct an in-depth
investigation of five popular metaverse platforms. Therein, we design and implement
\empirical{five} novel attacks that exploit creator-provided capabilities to
violate spatial, visual, and auditory constraints, with the ultimate goal of
achieving \emph{omniscience} within our world. These attacks enable a malicious
creator to covertly observe users, monitor private conversations, and manipulate
what users see and hear, without relying on software vulnerabilities or
developer-level privileges. We further show that \empirical{five} previously
proposed metaverse attacks can be replicated using only standard world-creation
tools, greatly lowering the barrier for adversarial behavior.

Our study shows that metaverse platforms introduce significant systemic
security and privacy risks by delegating extensive control over virtual worlds to
ordinary users. We show that world creators can carry out a broad range of
attacks, including covert surveillance and user monitoring,
using only standard
world-creation tools, without exploiting software vulnerabilities. These
attacks remain feasible across platforms and tooling models, as attackers can
adapt their strategies regardless of whether worlds are built using simplified
visual editors or advanced scripting interfaces. Moreover, we find that existing
platform defenses are largely ineffective: world vetting processes focus on
content rather than behavior, runtime permission systems and warnings are absent
or easily bypassed, and users receive no %
indication when
privacy-violating logic is present. Overall, our research demonstrates that any
user can readily assume the role of a world creator and deploy malicious
worlds at scale, while platforms currently lack meaningful safeguards to
protect users from the significant risks inherent to this model.

In summary, we make the following contributions:

\begin{itemize}
    \item Provide the first systematic security and privacy assessment of metaverse
    world creators, including a taxonomy of \empirical{25} platforms and their
    world-creation capabilities.
    \item Design and implement \empirical{five} novel attacks that exploit
    world-creator capabilities to violate spatial and sensory constraints,
    enabling user surveillance and manipulation.
    \item Show that \empirical{five} previously proposed metaverse attacks can be
    realized using only platform-provided world-creation tools, without software
    vulnerabilities or developer privileges.
    \item Evaluate platform vetting, runtime protections, and creator policies, and
    find them insufficient to mitigate malicious world-creator behavior.
\end{itemize}

\section{Threat Model}
\label{sec:threat_model}

We now define the threat model underlying our study. Our focus is on
privacy and security violations that arise from the legitimate design
capabilities delegated to metaverse world creators, rather than from software
vulnerabilities or platform misconfigurations. In particular, we study how a
malicious creator can violate users' expectations of spatial, visual, and
auditory privacy within a creator-controlled world.

\begin{figure}
    \centering
    \includegraphics[width=0.9\columnwidth]{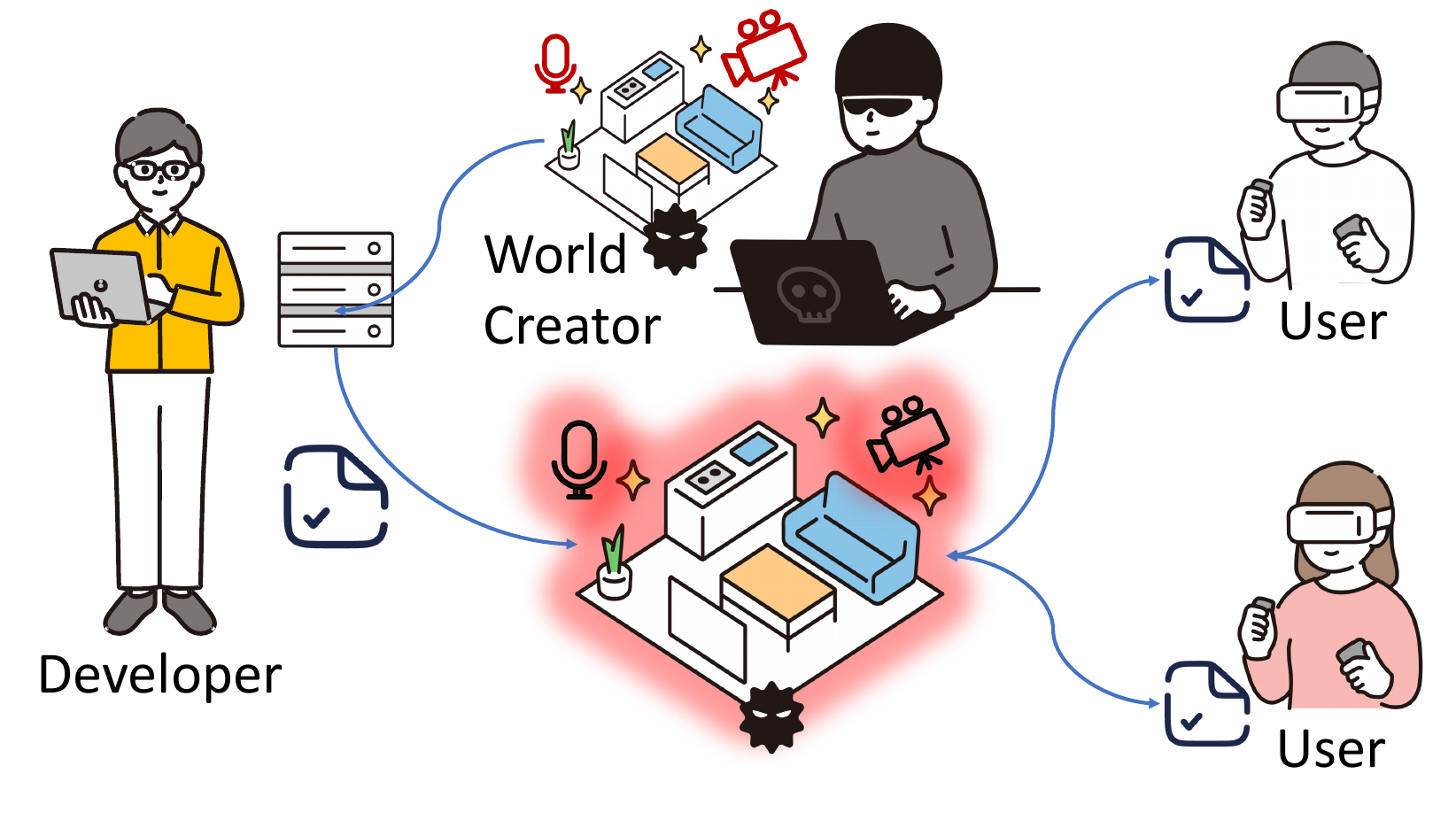}
    \Description{A developer publishes creation tools through the platform server. A malicious world creator uses them to embed virtual microphones and cameras into a world, which the platform then delivers to two unsuspecting users wearing VR headsets.}
    \caption{Threat model and actors in a metaverse platform.}
    \label{fig:threat_model}
\end{figure}
\textbf{Privacy expectations and existing trust models.} Our threat model
assumes benign users with privacy expectations shaped by two sources: the
immersive cues presented by the platform, and prior experience with other
multi-user communication systems.

Prior work shows that perceived enclosure in virtual
environments encourages self-disclosure, as users feel protected by virtual
boundaries such as walls or doors~\cite{sykownik2022something}. Metaverse
platforms reinforce these expectations through visual occlusion and spatial
audio, including distance-based attenuation and, recently, material-aware sound
propagation~\cite{BetaAcou26:online}. Users consequently treat these cues as
implicit guarantees of exclusivity: opaque barriers are assumed to block visual
observation, and conversations that fade with distance are presumed inaudible
to outsiders~\cite{limbago2025don,williamson2021proxemics,bailey2025students}.
Users actively rely on features such as privacy zones and lockable rooms as
safeguards, and report strong feelings of exposure and betrayal when these
spatial metaphors are silently
violated~\cite{mai2018feeling,bailey2025students}.
Crucially, users do not assume these guarantees as absolute: they understand
that a host may technically bypass them.  What they expect, however, is that
such access leaves a trace. Recent user studies found that even technically educated users who
acknowledge an administrator could override privacy features still expect this to
be signalled rather than silent, and treat covert listening as the main ethical
concern~\cite{bailey2025students}.

This expectation is not unique to immersive environments; it mirrors the trust
model of platforms that precede social metaverse systems, such as Slack,
Microsoft Teams, and Zoom. 
In each, administrators control membership and configuration,
but do not have default, invisible access to private conversations.  Microsoft
Teams' private channels are accessible only to their owners and members; even a
team administrator cannot access a private channel unless added as a
member~\cite{MicrosoftTeamsPrivateChannels}. Zoom hosts can join breakout
rooms, but they do so as visible participants rather than being able to silently monitor
all rooms at once~\cite{ZoomBreakoutRooms}. Chat history export, discovery, and
compliance mechanisms exist on some platforms, but they are treated as
exceptional and tightly
governed~\cite{SlackImportExportGuide,SlackExportWorkspace}. The lesson users
carry over is consistent: whoever organizes the interaction space is either
visible when present, or has to clear a high, auditable bar to access private
exchanges.

As such, it is evident that users find covert, unsignalled surveillance by whoever configured the space neither expectable nor acceptable.
Our threat model therefore assumes users expect no world
creators---ordinary users who configure and publish worlds---to have monitoring
capabilities beyond visible, in-world participation.

\textbf{Adversary: malicious world creators.}
We consider the adversary to be a malicious world creator, as illustrated in \Cref{fig:threat_model}. Any regular
platform user can become a world creator through a straightforward process,
typically requiring no additional identity verification beyond account
creation. For example, Horizon Worlds allows users to directly enter a creation
mode and publish worlds, while platforms such as Roblox and VRChat rely on
external editors that integrate seamlessly with user accounts. Although VRChat
requires users to meet activity-based ``Trust Rank'' criteria to publish public
worlds, this requirement remains accessible and does not meaningfully prevent
malicious participation.

We assume that the attacker can create multiple user accounts, use the official
creation tools to publish worlds in the metaverse platform catalogue, and
optionally join their own worlds as a regular participant. Similar to prior
work in other domains (e.g., analyzing phishing
pages~\cite{zhang2021crawlphish} or malicious browser
extensions~\cite{kapravelos2014hulk}), our focus is \emph{not} on how users are
lured into entering such worlds, but on what a creator-controlled world can do
once users join it.

\textbf{Attacker capabilities and action space.} 
The attacker's capabilities are strictly confined to those allowed by official
world-creation tools. 
Metaverse platforms provide editors, either client-integrated or standalone, that allow creators to design geometry, upload assets, configure materials, and define interaction logic via scripting or visual tools.
Such editors range from accessibility-oriented editors with limited logic primitives to advanced SDKs atop full-featured game engines (e.g., Unity with C\# scripting).

The malicious logic considered in this paper is implemented entirely using these
legitimate tools. The world creator does not exploit software vulnerabilities,
such as memory corruption, improper authorization, or unintended API access.
Creators do not have access to server-side databases, low-level networking
interfaces, or platform-wide state. Their influence is limited to the worlds and
instances they create or control.

For our empirical exploration of novel attacks, we assume an attacker whose
objective is \emph{covert omniscience}: the ability to covertly observe all users in a
world visually and aurally, regardless of spatial constraints that normally
limit perception, as we will see in \Cref{sec:new_attacks}. 
This is analogous to a communication-platform administrator being able to silently access private exchanges.
Beyond these novel attacks, the same creator capabilities
substantially lower the technical barrier for previously proposed attacks which
required client-side tampering or developer-level 
access. For our
re-implementation of these attacks, we adopt the objectives defined in prior
work (as detailed in \Cref{sec:editor-analysis}).

\section{World Creation Tools}
\label{sec:tools_attack_cat}

First, we survey and analyze the world-creation toolchains of popular
metaverse platforms. We (i) derive a classification of creator toolkits,
and use it to select representative platforms for our empirical evaluation, and then 
(ii) operationalize the primitives exposed to creators into a set of
technical capabilities that ground our novel attack designs (\Cref{sec:new_attacks}),
and our reproduction of prior attacks (\Cref{sec:editor-analysis}).

\begin{figure}
    \centering
    \includegraphics[width=0.8\linewidth, keepaspectratio]{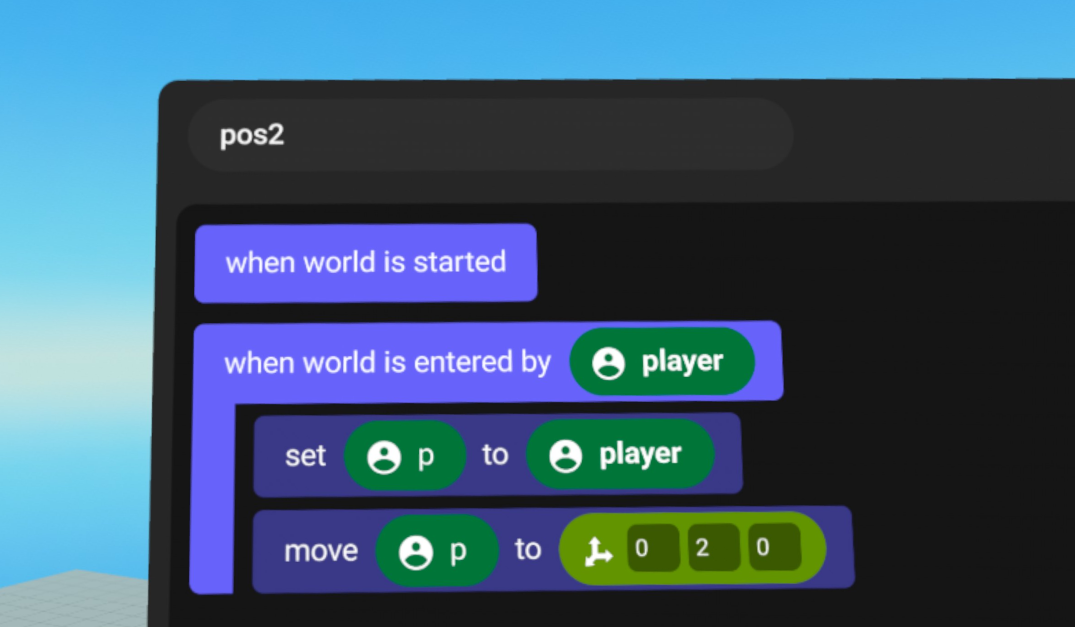}
    \Description{Screenshot of the block-based visual scripting editor in Horizon Worlds. The script reacts to world start and player entry events, stores the entering player in a variable, and moves the player by a fixed coordinate offset.}
    \caption{No-code editor (code blocks) in Horizon Worlds.}
    \label{fig:code_block}
\end{figure}

\subsection{Overview of Creation Tools}
\label{subsec:metaverse-platforms-survey}

We identified metaverse platforms that support user world creation via a survey
of search engines, VR app stores, and third-party databases. Using strict
inclusion criteria (VR support, multiplayer access, general availability, and
world-creation functionality), we compiled an initial set of \empirical{25}
candidates. Following common practice in web/mobile ecosystem
studies~\cite{257047,Zimmeck2016AutomatedAO}, we restricted the study to
platforms that are freely accessible to users. Two researchers independently
and iteratively reviewed developer documentation, tutorials, and user manuals for all \empirical{25} platforms, explored the creator tooling (apps/editors) of \empirical{19}, and fully categorized \empirical{14}, extracting a common set of
high-level creation primitives (editor type, logic expression, and
creator-facing APIs) and reconciling disagreements. The complete results of our
survey are in the \Cref{app:platform-selection} (\Cref{tab:metaverse_capabilities_with_discarded}),
whereas the summary of the platforms relevant for the security analysis of this
paper are in \Cref{tab:metaverse_capabilities_selected}.

\textbf{Editor models.} Across all platforms, we observe three recurring editor models. Some platforms build 
on established game engines by providing platform-specific SDKs (typically for
Unity), allowing creators to develop content externally and upload worlds
through platform tooling. Others integrate creation directly into the metaverse
client via in-app editors, enabling rapid in situ iteration. Finally, some
platforms offer custom external editors, standalone desktop or web-based
``studio'' tools, that provide a tailored workflow without relying on a
mainstream engine.

\textbf{Logic development.}
Orthogonal to the editor model, platforms differ in how creators express world
logic. Some rely on no-code visual interfaces exposing a fixed set of primitives
(e.g., block-based scripting in \Cref{fig:code_block}), while others provide
textual scripting or programming languages (e.g., C\#, Lua, or JavaScript),
offering greater expressiveness and direct support for complex behaviors.

\begin{table}[t]
\centering
\setlength{\tabcolsep}{3pt}
\renewcommand{\arraystretch}{1.15}
\resizebox{\columnwidth}{!}{
\begin{tabular}{l | ccc ccl | l}
\toprule
& \multicolumn{3}{c}{\textbf{Editor}} & \multicolumn{3}{c|}{\textbf{Logic}} &
\\
\cmidrule(lr){2-4} \cmidrule(lr){5-7}
\textbf{Metaverse} &
\rotatebox{90}{Game} & \rotatebox{90}{In-App} & \rotatebox{90}{Cust.} &
\rotatebox{90}{No-code} & \rotatebox{90}{Code} & \rotatebox{90}{Lang} & \textbf{Scale (metric + source)}
\\
\midrule

\textbf{Roblox}~\cite{HomeRobl54:online} &
& & \CIRCLE &
& \CIRCLE & Lua &
150M DAU, 3.5M creators~\cite{MetricsI65:online} \\

\textbf{Horizon Worlds}~\cite{MetaHori53:online} &
& \CIRCLE & \CIRCLE &
\CIRCLE & \CIRCLE & TS &
300k users (2022)~\cite{Metasso85:online} \\

\textbf{VRChat}~\cite{VRC:online} &
\CIRCLE & & &
\CIRCLE & \CIRCLE & C\# &
100k avg. concurrent users~\cite{Grafana35:online} \\

\textbf{Spatial}~\cite{SpatialC52:online} &
\CIRCLE & & &
& \CIRCLE & C\# &
1M+ creators~\cite{AboutSpa67:online} \\

\textbf{Frame}~\cite{Frame91:online} &
& \CIRCLE & \LEFTcircle &
\CIRCLE & \LEFTcircle & JS &
500k global users~\cite{FrameImm8:online} \\

\bottomrule
\end{tabular}
}
\caption{Creation toolchains for the \empirical{five} evaluated platforms, \protect\CIRCLE~if provided.
\protect\LEFTcircle: Frame dropped support for its Custom Editor (JS) after our experiments.
DAU: daily active users.}
\label{tab:metaverse_capabilities_selected}
\end{table}

\subsection{Creation Capabilities}
\label{subsec:capabilities}

Next, we identify features exposed to creators through platform-provided creation
tools. We focus on those that can be misused to violate users' privacy
expectations in virtual environments by subverting architectural or sensorial
cues (e.g.,~unauthorized and covert audio or video surveillance). We also identify the
technical primitives required to mount previously reported attacks from the
literature (see \Cref{sec:editor-analysis}).

A detailed analysis of all \empirical{25} surveyed platforms is infeasible, as
per-platform capability in-depth analysis is labor-intensive. We therefore select
\empirical{five} platforms that are representative of the surveyed set, covering the dominant combinations
of tool models and logic-coding paradigms (see
\Cref{tab:metaverse_capabilities_selected}). The full selection methodology,
inclusion and exclusion criteria, and the complete taxonomy are reported in
\Cref{app:platform-selection} and our open-science materials.

\textbf{1) Audio.} 
Platforms expose markedly different levels of control over users' audio streams.
At one extreme, Roblox enables flexible, programmable audio routing, allowing
creators to connect arbitrary audio inputs (e.g., user microphones) to arbitrary
outputs (e.g., in-world speakers), effectively decoupling speech from avatar
proximity. At the other extreme, Frame, Spatial, and Horizon Worlds strictly bind
audio to avatar position and distance, preventing creators from accessing,
redirecting, or otherwise manipulating audio beyond default spatial attenuation.
VRChat is in an intermediate position: audio remains tied to the speaker's
avatar, but creators can adjust parameters such as hearing range and volume,
enabling selective amplification or suppression of audibility.

\textbf{2) Video.} 
Access to virtual camera objects also spans a broad spectrum. Roblox provides
the most expressive model: creators can programmatically create and arbitrarily
position cameras, and stream their feeds to in-world displays for persistent
observation from any virtual viewpoint.
VRChat and Spatial support custom, movable cameras but require feeds to be
rendered onto visible in-world surfaces.
Horizon Worlds allows cameras to attach
to invisible, movable entities via scripting, but only on web and mobile clients
(not in VR). Frame adopts the most restrictive design: custom cameras are
unsupported, and creators are limited to either the fixed third-person browser camera or the first-person VR one 
that cannot be scripted or repositioned independently of the user's avatar.

\textbf{3) Custom objects.}
All five platforms allow creators to import custom 3D assets. At the time of our
study, all evaluated platforms supported scripting and allowed creators to modify
object visibility, behavior, and interaction logic at runtime, enabling reactive
environments. After our study, Frame dropped support for scripting, as discussed
in \Cref{subsec:limitations}.

\textbf{4) Network access.}
The ability to transmit data outside the platform enables attacks involving
logging, tracking, or exfiltration, and platforms impose widely varying
restrictions on this capability. Horizon Worlds disallows outbound network
requests entirely. VRChat permits limited HTTP GET requests to external
resources, restricted to allowlisted domains by default; users may optionally
enable an ``Allow Untrusted URLs'' setting, but POST requests are unsupported.
Spatial allows HTTP requests only in public worlds or with a \$100/month business
subscription and explicitly blocks requests to Spatial-owned domains.
Roblox disables HTTP requests by default, but creators can enable them via world
settings; requests are processed server-side and rate-limited (500~req/min).
At the time of our study, Frame is the most permissive: because worlds execute in the client browser,
creators can issue unrestricted HTTP GET and POST requests from client-side
scripts.

\textbf{5) Object attributes: read and write.} Modifying non-positional object attributes, such as texture, transparency, or
visibility, allows for customizing the design of the virtual environment.
All platform editors support reading and writing such attributes. Horizon Worlds
is a partial exception: its block-building editor lacks native support for
transparent or semi-transparent textures, though creators can achieve
transparency by adjusting alpha values (including negative values) or by using
the Desktop Editor.

\textbf{6) Spatial pose data (SPD): read and write.} 
All platforms maintain spatial pose data (position and orientation) for users
and objects. Reading and writing this data is universally supported across
creation tools, as it underpins core gameplay and interaction features.
Horizon Worlds is the sole exception: modifying user pose data requires enabling
a world-level setting, which the editor claims triggers a user warning (we
did not observe this in practice). Editing object pose data remains
universally supported.

\textbf{7) User identifiable information (UII).} 
Platforms differ in how persistently world creators can identify users across
sessions. Frame exposes usernames and, with explicit user consent, registration
email addresses. Roblox provides immutable user IDs and usernames, although
usernames can be changed for a fee ($\approx\$17$~USD). Horizon Worlds and VRChat
expose usernames but limit how frequently they can be changed (once every six
months and every 90 days, respectively). Spatial exposes the richest identifier
set, including immutable user IDs, usernames, and display names, with usernames
and display names freely editable by users.

\textbf{8) Continued execution.} 
Many attacks (e.g., long-term tracking, fingerprinting) require logic that
executes continuously over extended periods. All evaluated creation tools
support this through looping constructs or frame-based
callbacks, enabling creators to deploy scripts that run indefinitely within
their worlds.

\subsection{Takeaways} 
We surveyed \empirical{25} metaverse platforms and distilled their
world-creation tools into a set of recurring design patterns, which we
instantiate through \empirical{five} representative platforms. Across these,
world-creation tools expose a rich and largely uniform set of low-level
primitives (e.g., persistent execution, access to spatial state, object and attribute
control, and user identifiers) that are sufficient to assemble privacy-invasive
behaviors. While only a few actions are categorically forbidden, most safeguards
constrain specific implementations rather than underlying capabilities, leaving
semantically equivalent alternatives available, as we demonstrate in
\Cref{sec:new_attacks} and \Cref{sec:editor-analysis} by presenting concrete
attacks. As a result, these restrictions act primarily as friction: by recombining
legitimate features, creators can systematically undermine users' spatial and
sensory privacy expectations without exploiting vulnerabilities or requiring
developer privileges. These primitives form a strict subset of those available to platform developers: creators cannot reach the OS or device-level APIs that enable, e.g., side-channel or keystroke attacks, but they inherit the application's multi-user network model, which exposes the state of all co-present users. However, as we will see, this narrower toolset already suffices to achieve omniscience. The following sections concretely instantiate these
building blocks into both novel and previously-proposed attacks.

\section{Omniscience Attacks}
\label{sec:new_attacks}

\begin{table*}[h!]
\small
\centering

\setlength{\tabcolsep}{3pt}
\resizebox{\textwidth}{!}{
\begin{tabular}{l | c c c c c c c c | c c c c c | c | l}
\toprule
 & \multicolumn{8}{c}{\textbf{Capabilities}} & \multicolumn{5}{c}{\textbf{Platform}} & & \\
\cmidrule(lr){2-9} \cmidrule(lr){10-14}
\textbf{Attack} &
\rotatebox{90}{Audio} &
\rotatebox{90}{Video} &
\rotatebox{90}{Cust.\ Obj.} &
\rotatebox{90}{Obj.\ Attr.} &
\rotatebox{90}{Network} &
\rotatebox{90}{SPD} &
\rotatebox{90}{UII} &
\rotatebox{90}{Cnt'd Exec.} &
\rotatebox{90}{VRChat} &
\rotatebox{90}{Roblox} &
\rotatebox{90}{Spatial} &
\rotatebox{90}{Frame} &
\rotatebox{90}{Worlds} &
\rotatebox{90}{\textbf{Automated}} &
\textbf{Impact} \\
\midrule

\multicolumn{16}{l}{\emph{Omniscience Attacks}} \\
\hspace{1em}Parabolic Microphone         
    & \ding{72} & ~ & ~ & \ding{72} & ~ & \ding{72} & ~ & ~
    & \CIRCLE & \CIRCLE & ~ & ~ & ~ 
    & \fullDiamond
    & Monitors audio of remote users' conversations. \\
\hspace{1em}Control Room                 
    & ~ & \ding{72} & ~ & \ding{72} & ~ & \ding{72} & ~ & ~
    & \CIRCLE & \CIRCLE & \CIRCLE & ~ & ~ 
    & \fullDiamond
    & Monitors video across multiple private locations. \\
\hspace{1em}Astral Projection            
    & ~ & \ding{72} & ~ & ~ & ~ & \ding{72} & ~ & ~
    & \CIRCLE & \CIRCLE & \CIRCLE & \CIRCLE & \CIRCLE 
    & \halfDiamond
    & Monitors spaces via attacker-controlled invisible camera. \\
\hspace{1em}Unidirectional Material 
    & ~ & ~ & \ding{72} & \ding{72} & ~ & ~ & ~ & ~
    & \CIRCLE & \CIRCLE & \CIRCLE & \CIRCLE & \CIRCLE 
    & \halfDiamond
    & Monitors users through one-way walls or floors. \\
\hspace{1em}Conversation Hijacking       
    & \ding{72} & ~ & ~ & ~ & ~ & ~ & \ding{72} & ~
    & \CIRCLE & \CIRCLE & ~ & ~ & ~ 
    & ~
    & Replaces target user's audio to manipulate conversations. \\

\midrule
\multicolumn{16}{l}{\emph{State-of-the-Art Attacks}} \\
\hspace{1em}User Re-identification~\cite{Munilla_Garrido_Nair_Song_2024}
    & ~ & ~ & ~ & ~ & \ding{72} & \ding{72} & \ding{72} & ~
    & \LEFTcircle & \CIRCLE & \LEFTcircle & \CIRCLE & \LEFTcircle 
    & \fullDiamond
    & Tracks users persistently across sessions or worlds.\\
\hspace{1em}Social Logging~\cite{Vondrek_Baggili_Casey_Mekni_2023}
    & ~ & ~ &  & ~ & \ding{72} & \ding{72} & \ding{72} & \ding{72}
    & \LEFTcircle & \CIRCLE & \LEFTcircle & \CIRCLE & \LEFTcircle 
    & \fullDiamond
    & Profiles users by recording interactions and behaviors.\\
\hspace{1em}User DoS~\cite{Mengascini_Aurelio_Pellegrino_2024}
    & ~ & ~ &  & ~ & ~ & \ding{72} & \ding{72} & \ding{72}
    & \CIRCLE & \CIRCLE & ~ & \invdiameter & \LEFTcircle 
    & \fullDiamond
    & Disrupts user participation by blocking or isolating them.\\
\hspace{1em}FP with Motion Data~\cite{Nair_Rack_Guo_Wang_Li_Huang_Cull_OBrien_Latoschik_Rosenberg_et}
    & ~ & ~ &  & ~ & \ding{72} & \ding{72} & \ding{72} & \ding{72}
    & \LEFTcircle & \CIRCLE & \LEFTcircle & \invdiameter & \CIRCLE 
    & \fullDiamond
    & Tracks and profiles users based on movement patterns.\\
\hspace{1em}Human Joystick~\cite{Casey_Baggili_Yarramreddy_2021}
    & ~ & ~ &  & ~ & ~ & \ding{72} & \ding{72} & \ding{72}
    & \CIRCLE & \LEFTcircle & \CIRCLE & \invdiameter & \LEFTcircle 
    & \fullDiamond
    & Manipulates user's movement or view covertly.\\

\bottomrule
\end{tabular}
}
\caption{
Required capabilities and feasibility of each attack across metaverse platforms.
\ding{72}: Capability required for the attack; \CIRCLE: Feasible with platform capabilities; 
\LEFTcircle: Feasible with alternative implementation; \invdiameter: Not feasible in Frame anymore, due to scripting support being discontinued after the completion of our experiments. \protect\fullDiamond: Automated; \protect\halfDiamond: Automated under conditions.
}
\label{tab:attacks_platforms_impact}
\end{table*}

We now present novel attacks that demonstrate the inherent threat posed by the
world creator role emerging in metaverse platforms. We focus on devising new
techniques that allow the attacker to achieve omniscience within the worlds
they create, by reaching a state of total audio and video surveillance without
the introduction of any auditory or visual artifacts that could potentially
alert users about an ongoing attack. At heart, our attacks build on the audio
and video capabilities offered by the world creator tools, focusing on two
separate modalities:

\begin{enumerate}%
\item \textbf{Superhuman senses.} As the world creator, the attacker has
enhanced capabilities regarding the manipulation of their
own, or others', audio and video signals.

\item \textbf{Perceptual illusions.} World creation allows the attacker 
to create environmental illusions that target specific users or widely
apply to all other users, and manipulate their perception of %
the surrounding environment and its properties.
\end{enumerate}

In the following subsections we present our attack techniques, and provide
additional information about their internal workings and their implementation
in specific metaverse platforms. The attacks are summarized in Table~\ref{tab:attacks_platforms_impact}.
Anonymized video demonstrations of all attacks~\cite{Metavers85:online} are available in \Cref{sec:openscience_datasharing},
with textual explanations and timestamps for each phase.

\textbf{Attack design.} We note that our attacks \emph{do not} require the
presence of vulnerabilities, nor do we attempt to uncover any during our
empirical analysis. We design our novel attacks using the creator tools as
intended; in cases where we do not have direct access to a specific data
source, we explore whether we can achieve the objective by
designing ``malicious'' world logic.
Ultimately, our attacks demonstrate that 
\emph{creative} usage of world creation tools allows for the deployment of
powerful attacks with significant privacy implications for users.
Moreover, we note that while the concrete implementation of each 
attack can vary across platforms depending on the APIs, objects, and 
scripting capabilities they provide, the overall design remains consistent.
As such, our attacks can be straightforwardly implemented in other platforms.

\textbf{Experimental setup.} We created three accounts in each of the
\empirical{five} metaverse platforms, one as creator and the others as
victims. Since we had two Oculus Quest devices (Oculus
Quest 2 and 3), the third account entered the environment either through the web
or mobile client, depending on platform support. We selected the attacker's
client (VR or non-VR) based on which option maximized the success and ease of
the attacks, using specific clients whenever they exposed additional
information or functionalities. All attacks were developed and executed using
a Lenovo ThinkPad P1 Gen 6 with an Intel i9-13900H with 32 GB of RAM and a GPU
RTX4090 mobile running Windows 11 Pro. We chose the platforms' option to build
and compile a world for multiple platforms, e.g., web, mobile, and VR. All
attacks were evaluated in a controlled laboratory setting and underwent
multi-researcher validation to ensure correctness and reproducibility. We note that 
for all of our attacks, when audio surveillance is involved users' speech is captured
without any noticeable loss or deterioration, and video monitoring clearly
captures user avatars and actions.

\subsection{Parabolic Microphone}
\label{sec:parabolic_microphone}
\begin{figure}
    \centering
    \includegraphics[width=0.7\columnwidth, keepaspectratio]{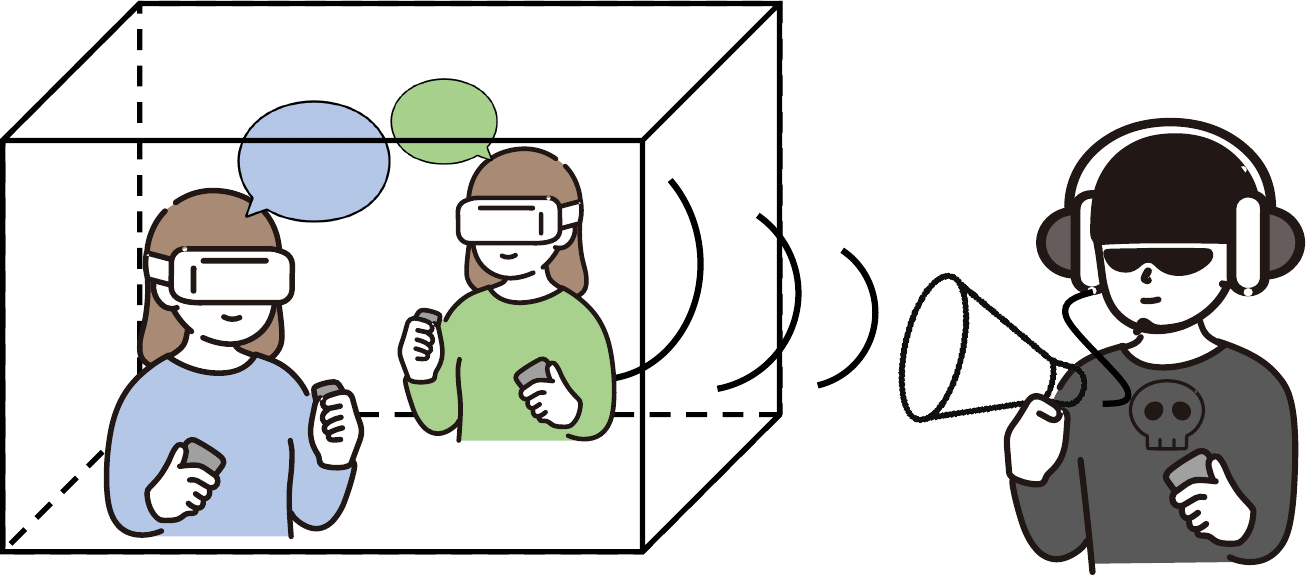}
    \Description{Two users converse inside a virtual room while an attacker, positioned outside the room with headphones and a listening horn, captures their speech from a distance at which they would normally be inaudible.}
    \caption{The parabolic microphone attack.}
    \label{fig:parabolic_mic}
\end{figure}
This targeted eavesdropping attack (\Cref{fig:parabolic_mic}) allows an attacker to listen to
conversations within a metaverse world, regardless of spatial distance or the
presence of physical barriers (e.g., walls). The attack relies on audio
manipulation capabilities available only to world creators, through
platform-provided APIs, to alter how sound is transmitted and perceived.
Specifically, the attacker can amplify or re-route audio from users,
making conversations audible only to them, without affecting how other
participants experience sound in the environment.

\textbf{VRChat.} This platform provides audio functions to the creator
that can be used to modify how they hear other users in the world. We used two
functions to implement our attack: \texttt{VRCPlayerApi.SetVoiceGain} and
\texttt{VRCPlayerApi.SetVoiceDistanceFar}. The first function adjusts a
specific player's volume, and the second sets the maximum distance for hearing
that player's voice; these are executed on the
client side. The script runs only when the local player's name matches the
attacker's name, ensuring that the changes only affect how the attacker hears
others.

Using the provided functions we can choose which players we want to hear. We
created a script that first checks whether the local player is the attacker. If
that is the case, we set \texttt{VRCPlayerApi.SetVoiceDistanceFar} to the
maximum value for all players, allowing the attacker to hear everyone from any
distance. Then, we created a UI object inside the world containing a list of
players and a slider for controlling each player's volume. We also attached the
slider to a script that uses \texttt{VRCPlayerApi.SetVoiceGain} based on the
value of the slider. Using this method, the attacker can choose which players
to hear by adjusting the voice sliders accordingly, allowing them to focus on
specific conversations.

\textbf{Roblox.} The Roblox Audio API enables developers to route and
manipulate audio using modular components such as \lstinline|AudioListener| (microphone), \lstinline|AudioEmitter| (speaker), and \lstinline|Wire| to
connect them. Leveraging these primitives, we implemented a parabolic
microphone attack that simulates a wiretapped environment. Our setup consists
of hidden ``microphone rooms'' containing an \lstinline|AudioListener| object, and
a separate ``speaker room'' accessible only to the attacker, containing an
\lstinline|AudioEmitter|. By linking these components with a \lstinline|Wire|,
we reroute real-time voice data to the attacker's location, bypassing audio
attenuation and occlusion due to environmental aspects (e.g., walls or distance). This
acts 
as a directional microphone, capturing conversations remotely
and replaying them elsewhere, 
undetectable by victims or other users.

\textbf{Other platforms.} The attack could not be replicated on the other
platforms due to the lack of appropriate audio APIs.

\subsection{Control Room}
\label{sec:control_room}
\begin{figure}[htbp]
    \centering
    \includegraphics[width=0.7\columnwidth, keepaspectratio]{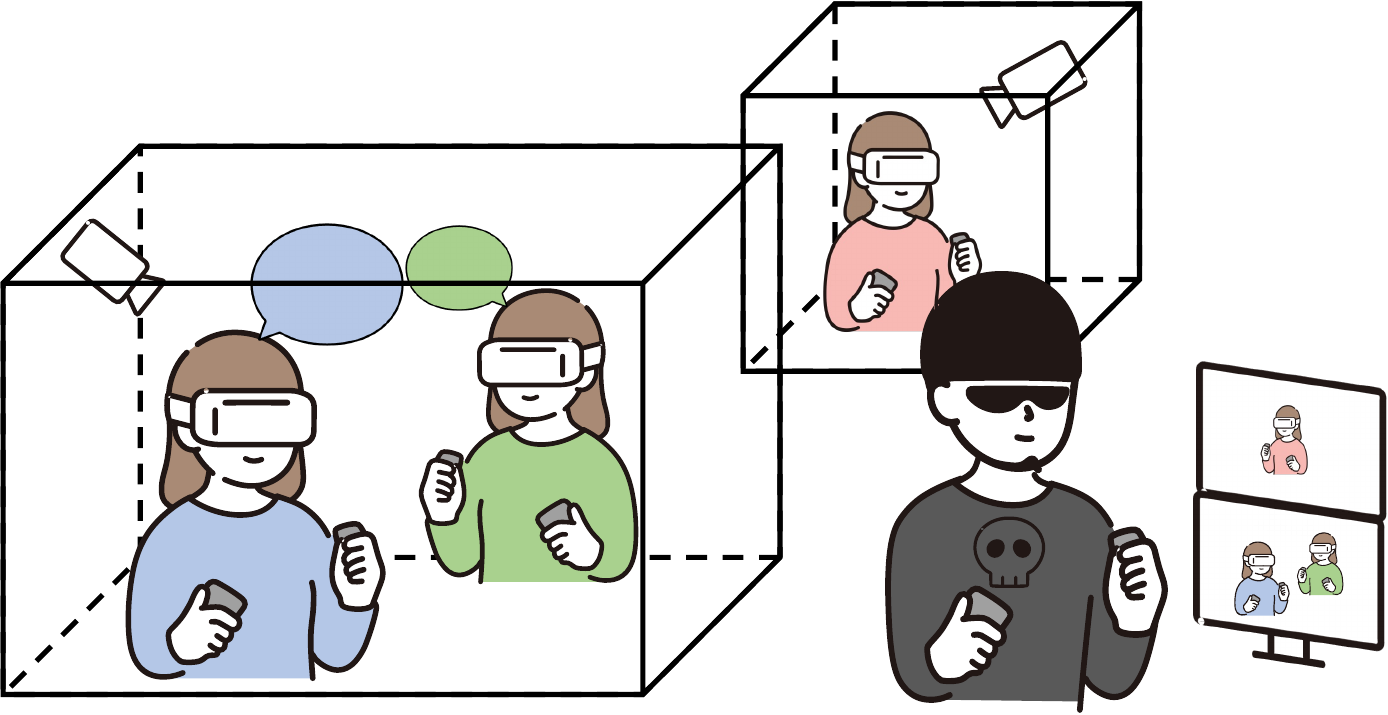}
    \Description{Camera objects placed in the corners of two separate virtual rooms stream their views to an attacker, who observes the occupants of both rooms on external monitors without being present in either room.}
    \caption{Control Room for surveillance.}
    \label{fig:control_room}
\end{figure}
\noindent Audio eavesdropping attacks capture speech but miss the visual
interactions central to metaverse experiences, such as gestures, movement, and
object use. To observe this activity, we developed a video-based
surveillance attack (\Cref{fig:control_room}) that allows passive monitoring across multiple regions of
the world. In this attack, the adversary creates a private
``control room'' containing a wall of screens, each displaying the live feed from
invisible cameras placed throughout the world. The cameras are strategically
positioned in common or sensitive areas to monitor user behavior without their
knowledge (e.g., board room meeting). To execute the attack, the platform
must support the creation of camera objects and the ability to redirect their
visual output to in-world display surfaces.

\textbf{VRChat and Spatial.}
Both platforms allow an attacker to place multiple \lstinline{Camera} objects
throughout the virtual world and display each camera's view on a
\lstinline{Plane} object that acts as a monitor. The \lstinline{Camera} objects
are invisible by default, allowing the attacker to observe users without their
knowledge. In VRChat, this video monitoring can be combined with parabolic
microphones to also capture conversations, providing a comprehensive view of
user activity.

\textbf{Roblox.}
Roblox does not provide a direct way to stream a camera's view onto a screen.
Instead, the attacker can use a virtual \lstinline{Camera} with a
\lstinline{ViewportFrame}---a UI object that renders 3D objects inside its
bounds~\cite{Viewport58:online}---to simulate CCTV surveillance. Invisible
cameras are placed in key areas, and a script cycles through all objects each
frame to determine what should appear in the camera's view, then renders this
to the screen. The attacker can view these feeds in a private control room or
an attacker-only visible HUD using a \lstinline{ScreenGui}. Combined with a
parabolic microphone, this enables real-time video and audio monitoring across
the world.

\textbf{Other platforms.} This attack is not feasible on platforms
that do not expose camera and screen-like objects to world creators, or that
restrict the ability to capture and replicate user or object movement within
the scene.

\subsection{Astral Projection}
\label{sec:astral_projection}
\begin{figure}[htbp]
    \centering
    \includegraphics[width=0.7\columnwidth, keepaspectratio]{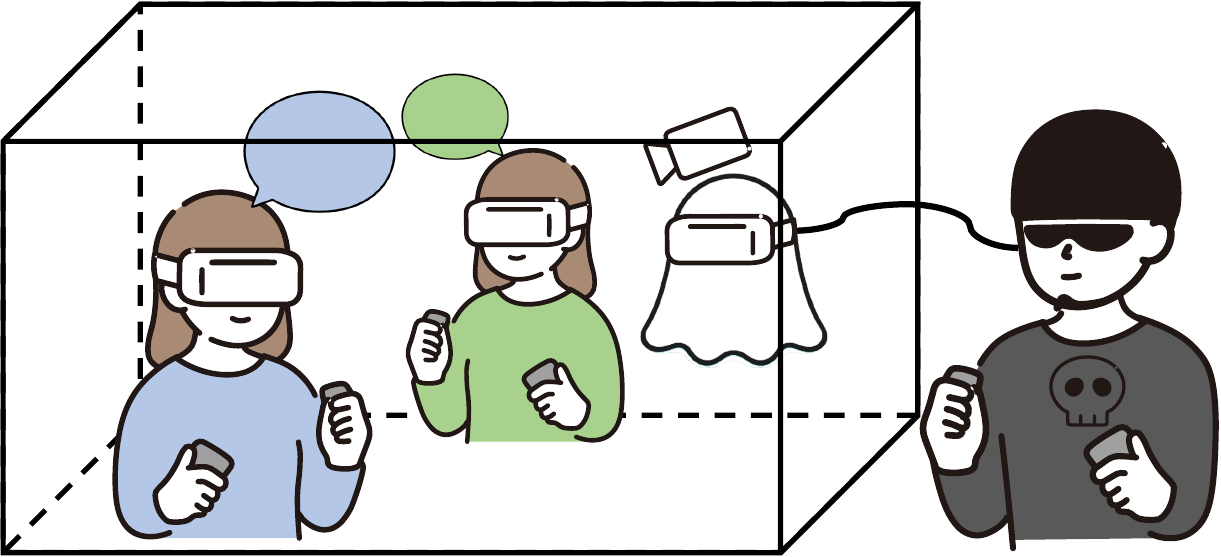}
    \Description{An attacker outside a virtual room steers an invisible, ghost-like entity carrying a camera that stands beside two conversing users, relaying their conversation and surroundings back to the attacker.}
    \caption{The astral projection attack.}
    \label{fig:astral_projection}
\end{figure}
\noindent This attack allows the attacker to decouple their viewpoint from their avatar,
effectively rendering them invisible to others while freely navigating their
world (\Cref{fig:astral_projection}). The attacker controls a virtual camera
independent of their in-world position, enabling them to monitor private
spaces and interactions. The attack requirements are lower
than those of the control room attack, as it does not require in-world screens,
only to move the attacker's own camera away from its avatar.
\Cref{alg:astral_projection} details the implementation. $A$ is the world
creator who later joins as a participant; upon joining, $A$ gains access
to the live player list for ID-based target selection, then decouples the camera
to surveil $V$ via key input or automatic position tracking.

\begin{algorithm}[htbp]
\caption{Astral Projection. $A$ (world creator) 
moves their camera near $V$, observing $V$ from any angle while invisible to all users.}

\label{alg:astral_projection}
\begin{algorithmic}[1]

\Statex \textbf{On world join} \textit{($A$ only, platform-specific)}:
\State \textbf{Roblox:}\ $\mathit{cam.CameraType} \gets \textsc{Scriptable}$ \hfill \pseudocomment{hijack player cam}
\State \textbf{VRChat\,/\,Spatial:}\ $\mathit{cam} \gets \Call{Camera}{\mathit{pos}}$ \hfill \pseudocomment{invisible world obj}
\State \textbf{Horizon:}\ $\mathit{cam} \gets \Call{Camera}{\mathit{pos}}$;\ $\Call{cam.attach}{\mathit{entity}}$

\Statex
\Statex \textbf{Attacker selects target} $V$ \textit{(ID-based)}:
\State $V \gets \Call{GetPlayerByID}{\mathit{target\_id}}$ \hfill \pseudocomment{$V$ unaware of selection}
\State $\Call{camera.position.set}{V.\mathit{position}}$ \hfill \pseudocomment{camera to $V$'s location}

\Statex
\Statex \textbf{Attacker observes} $V$ \textit{(covert, persistent)}:
\While{$\mathit{true}$}
    \State $\Delta \gets \mathit{manual}\ ?\ \Call{ReadInput}{}\ :\ V.\mathit{position} - \mathit{cam.position}$
    \State $\Call{camera.position.set}{\mathit{cam.position} + \Delta}$
\EndWhile

\end{algorithmic}
\end{algorithm}
\textbf{Roblox.} The \lstinline|CameraType| property allows full programmatic
control of the player camera. By setting the camera to \lstinline|Scriptable|,
it can be moved throughout the environment using custom controls, regardless of
the avatar's position. The attacker's audio input also follows the camera, so
they can hear conversations from the camera's current location, enabling both
visual and audio surveillance.

\textbf{VRChat and Spatial.} Here it is
not possible to independently move the actual player camera. Instead, the
attacker creates a new camera object in the world, which is attached to a
screen-like object that only the attacker can see. The attacker can move and
rotate this camera anywhere in the environment with keyboard or controller
inputs (\lstinline|Input.GetKey|, \lstinline|transform.Translate|). This
allows for visual surveillance, but the camera does not
capture audio. To monitor conversations, the attacker must combine this attack
with an audio-based technique, such as the parabolic microphone.

\textbf{Horizon Worlds.} Creators can script cameras by
attaching them to invisible, movable objects, using the ``Fixed Camera Position
with Entity'' option via the \lstinline|playercamera| entity. Camera movement
is controlled through custom input buttons that reposition the underlying
object, enabling the attacker to navigate the environment independently of
their avatar. To use this feature, the attacker must connect through either
the web or the mobile client, as VR clients remain limited to first-person
view~\cite{MetaDeve73:online}. While this attack allows full visual
surveillance, it does not capture spatial audio; audio monitoring requires
combining a separate technique.

\textbf{FrameVR.} FrameVR does not support custom or scriptable
cameras. However, users accessing the world via a web browser in third-person
mode can zoom out and pan the camera nearly without restriction, observing most
of the environment from a distance. The camera orientation remains fixed on the
avatar and cannot be rotated freely, so the attacker can see who is present and
what they are doing, but not always from all angles. This technique provides
only visual monitoring; to eavesdrop on conversations, attackers would need to
combine this with an audio-based attack.

\subsection{Unidirectionally Transparent Material}
\label{sec:one_way_material}
\begin{figure}[htbp]
    \centering
    \includegraphics[width=0.7\columnwidth, keepaspectratio]{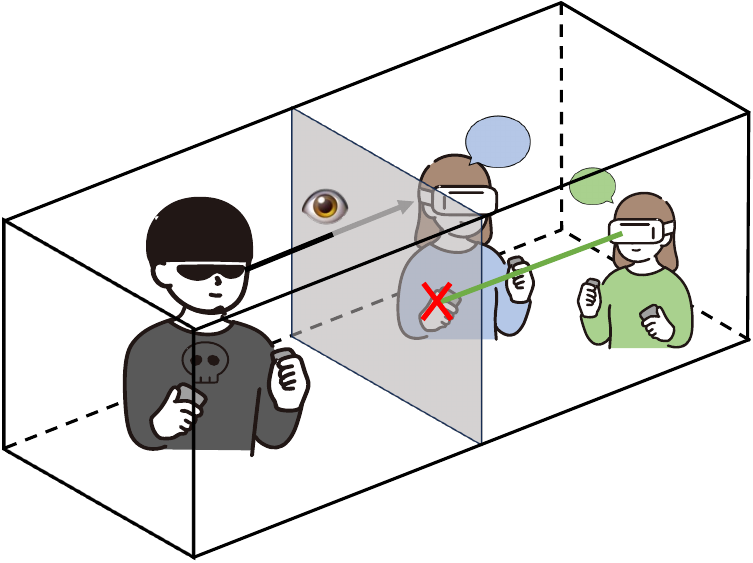}
    \Description{A barrier that is transparent in one direction only separates an attacker from two users: the attacker's line of sight passes through the barrier towards the nearest user, whereas the users' view of the attacker is blocked.}
    \caption{Unidirectionally Transparent Material.}
    \label{fig:one_way_material}
\end{figure}
\noindent This attack employs the perceptual illusion modality to build an environment
that creates the illusion of a private environment for users (\Cref{fig:one_way_material}). In more detail,
this attack relies on a carefully designed space inspired by the properties of
interrogation rooms (as seen in movies). Essentially, the attacker can employ
a variety of different designs to create a complex of rooms, where parts of the
room (i.e., floor, walls, ceiling) are comprised of a material that is one-way
see-through. This allows the attacker to place themselves behind a wall, or
under the floor, or on the roof, of a room designed to spy on users. Using a
unidirectionally transparent object, the attacker can see everything in the
victim's room through this object. However, other users perceive that object
as solid and therefore can not see what is behind it. The attacker can also
hear the players in the room if they are within hearing range. As the world
creator, the attacker can determine the room size, ensuring it falls within
their hearing range. Alternatively, the attacker can combine this attack with
any of the audio-based attacks. %
The only requirement for this attack is for the platform to support the
creation of custom objects.

\textbf{All platforms.} Since all platforms in our study
allow uploading custom 3D assets, we created a one-way see-through object using
Blender~\cite{blendero82:online}. To achieve this effect, we enabled the
\texttt{Backface Culling} option in the material shader settings and selected
the \texttt{Camera} flag. This configuration makes the object renderable from
only one side, appearing opaque on one side while 
transparent from the
other, 
creating a unidirectional window. We used this to
construct spy rooms that visually deceive users in all platforms.

\textbf{Platforms with scripting support.} Additionally, on
platforms that support user scripting (i.e., VRChat, Roblox, Spatial, Horizon
Worlds), we also implemented an ad-hoc, identity-based wall-visibility control
mechanism. We attach a script to the wall object that checks the local user's
username at runtime. By default, the object is enabled, but if the script
detects the attacker's name, it disables the object for them. For all other
users the wall appears solid, while for the attacker it is invisible, allowing
direct observation without needing custom shaders. This approach can complement
or replace custom object-based transparency.

\subsection{Conversation Hijacking}
\label{sec:conversation_hijacking}
\begin{figure}[htbp]
    \centering
    \includegraphics[width=0.7\columnwidth, keepaspectratio]{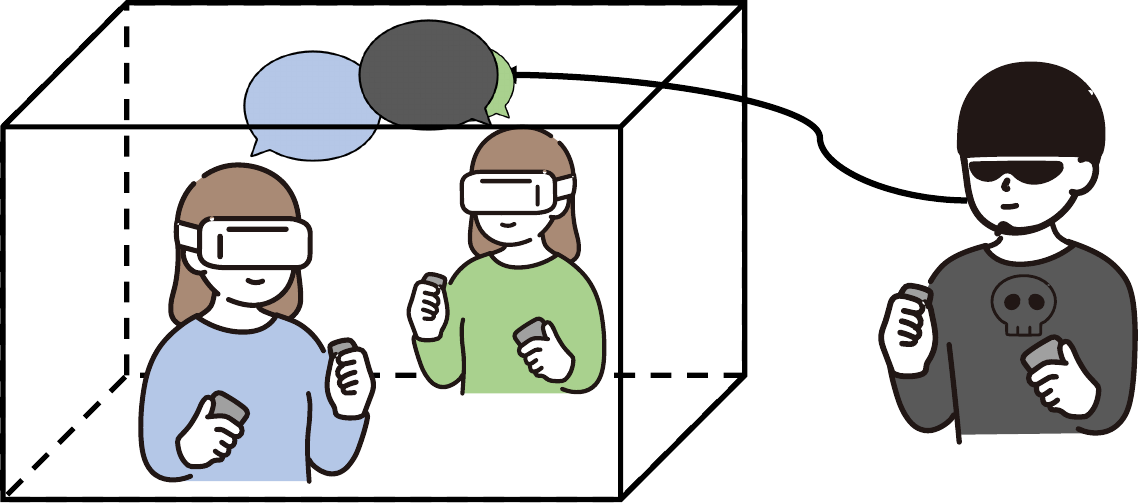}
    \Description{Two users converse inside a virtual room while an attacker outside the room injects a fabricated utterance, shown as a dark speech bubble inserted between their genuine ones, altering the conversation the users perceive.}
    \caption{The conversation hijacking attack.}
    \label{fig:conversation_hijacking}
\end{figure}
\noindent While the previous attacks are passive and focus on monitoring
users, we have also designed an attack that allows the attacker
to actively interfere with ongoing user interactions. This enables a
wide range of attack scenarios, including spreading misinformation,
malvertising, influencing the outcome of a collaborative decision, or
manipulating users' opinions. Specifically, the attacker impersonates another
user by muting the target player's audio and increasing their own volume
for the target's conversation partners. In other words, the attacker is
able to replace the target users' audio feed with an audio stream of their
choice (\Cref{fig:conversation_hijacking}). This attack requires
the ability to modify other players' audio settings.

To strengthen the attack, the adversary can pre-record
messages using a cloned version of the victim's voice or leverage near-instant
voice cloning tools for real-time impersonation. Recent work shows that
convincing voice clones can be generated from just a few seconds of
audio~\cite{qin2024openvoiceversatileinstantvoice}, making such attacks
increasingly practical.

\hspace*{\parindent}\textbf{VRChat.} Consider user A (victim) and user B (conversation
partner). The attacker uses a script to adjust user B's hearing of nearby
players. First, the attacker mutes user A by setting their audio distance to
zero for user B, using \lstinline|VRCPlayerApi.SetVoiceDistanceFar|. Then, the
attacker boosts their own volume to the maximum, making their voice heard by
user B. As a result, user B only hears the attacker, believing they are still
communicating with user A.
\Cref{alg:conv_hijack} details the implementation of the conversation hijacking attack for VRChat, where $V$ and other users are unaware of the manipulation.

\textbf{Roblox.} Here, the attacker achieves a similar effect by modifying
the \lstinline|Muted| property of user A's \lstinline|AudioDeviceInput| (microphone) object,
effectively silencing them. Simultaneously, the attacker uses an
\lstinline|AudioEmitter| (speaker) to inject their own voice into the environment,
targeting user B. To preserve the illusion of authenticity, the attacker can
script a fake “speaking” icon to appear next to user A's name, mimicking the
platform's native UI indicator for speaking. This visual manipulation further
deceives the victim into believing the impersonation is genuine.

\textbf{Other platforms.} This attack is not possible on platforms that do not
allow the world creator to modify other users' audio settings.

\begin{algorithm}[]
\caption{Conversation Hijacking (VRChat). $A$ silences $V$ for all users and
injects their own voice into $B$'s feed, making $B$ believe they are still
talking to $V$. $A$ then restores the original audio state.}
\label{alg:conv_hijack}
\begin{algorithmic}[1]

\Statex \textbf{On world start} \textit{(attacker only)}:
\State $\mathit{players} \gets$ \Call{GetAllPlayers}{}
\State \textbf{for each} $p \in \mathit{players}$: add entry with toggles \textsc{Mute}($p$), \textsc{Relay}($p$)
\State Update panel on player join / leave

\Statex
\Statex \textbf{Attacker activates hijack} (selects victim $V$, bystander $B$):
\State \Call{SetVoiceDistanceFar}{$V,\ 0$} \hfill \pseudocomment{silence $V$ globally}
\State Sync \textsc{Mute}($V$) and \textsc{Relay}($B$) state to all clients
\If{localPlayer $= B$} \hfill \pseudocomment{change only $B$'s state}
    \State \Call{SetVoiceDistanceFar}{$A,\ \infty$} \hfill \pseudocomment{$B$ hears $A$ instead of $V$}
\EndIf

\end{algorithmic}
\end{algorithm}

\subsection{Attack Automation and Stealthiness}
Next, we characterize our attacks by their degree of automation, i.e., whether they
can persist as world logic after deployment or require continued attacker
intervention. This characterization complements our 
platform feasibility 
analysis, summarized in \Cref{tab:attacks_platforms_impact}.

\emph{Parabolic Microphone} and \emph{Control Room} are automated across all
platforms where we implemented them: once configured, their logic is embedded
in the world and enables unattended A/V surveillance.

Other attacks expose a spectrum of automation depending on whether the
attacker 
monitors fixed locations or dynamically follows specific users.
\emph{Astral Projection} and \emph{Unidirectionally
Transparent Material} can be fully automated when monitoring static areas
(e.g., a private room), by placing a camera or the creator's viewpoint at a
predetermined position.
\emph{Unidirectionally Transparent Material} remains automatable in targeted scenarios by positioning the creator avatar in a concealed vantage point (e.g., below a one-way floor) and tracking the victim.
Similarly,
\emph{Astral Projection} supports target-following automation on all platforms
except FrameVR, where the camera cannot be locked onto a user and must
be manually repositioned. In both attacks, however, when the attacker wishes
to switch targets to adapt to unfolding context, manual intervention 
is needed, 
making both partially automated.

For automated and conditionally automated attacks, data collection is
straightforward using standard client-side screen and audio recording tools;
we use OBS~\cite{OpenBroa75:online} for web clients, and built-in recording on Meta Quest or mobile
devices.

Finally, \emph{Conversation Hijacking} is not automatable in practice. The
attack requires real-time semantic understanding and precise timing of audio
manipulation to impersonate a specific user in an ongoing interaction,
which current platforms do not support programmatically, leaving such attacks
reliant on human intervention.

\textbf{Stealthiness.} Our attacks introduce no auditory or visual artifacts
that would alert users to ongoing surveillance or interference. In surveillance
attacks, audio remains spatialized normally, and no visible indicators reveal
attacker-controlled cameras, concealed observation points, or rerouted audio
paths. 
Similarly, \emph{Conversation Hijacking} alters how select users
perceive a conversation without observable cues for the impersonated victim or
surrounding participants. Extensive testing confirmed that victim accounts
perceive no differences during attack execution compared to benign worlds.
We measured the client-side overhead our attacks impose on the victim, repeating each measurement three times per platform, except on Spatial, which dropped support during our study. The impact is negligible: on average, frame rate changes by $-1.0\%$ and GPU utilization by $-1.5\%$, while CPU utilization rises by only ${\sim}10\%$. We report the full per-attack, per-platform breakdown in \Cref{tab:overhead}.

\section{Revisiting State-of-the-Art Attacks}
\label{sec:editor-analysis}
\begin{figure*}[htbp]
    \centering
    \begin{subfigure}[b]{0.19\textwidth}
        \centering
        \includegraphics[width=\linewidth,height=2.8cm,keepaspectratio]{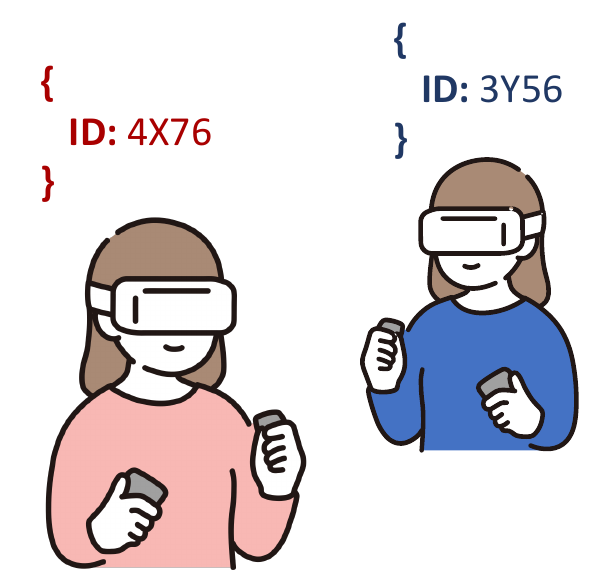}
        \Description{Two avatars, each annotated with a record containing a distinct platform-assigned identifier, illustrating that such identifiers allow users to be recognized across encounters.}
        \caption{Re-identification.}
        \label{fig:attack-fingerprinting}
    \end{subfigure}\hfill
    \begin{subfigure}[b]{0.19\textwidth}
        \centering
        \includegraphics[width=\linewidth,height=2.8cm,keepaspectratio]{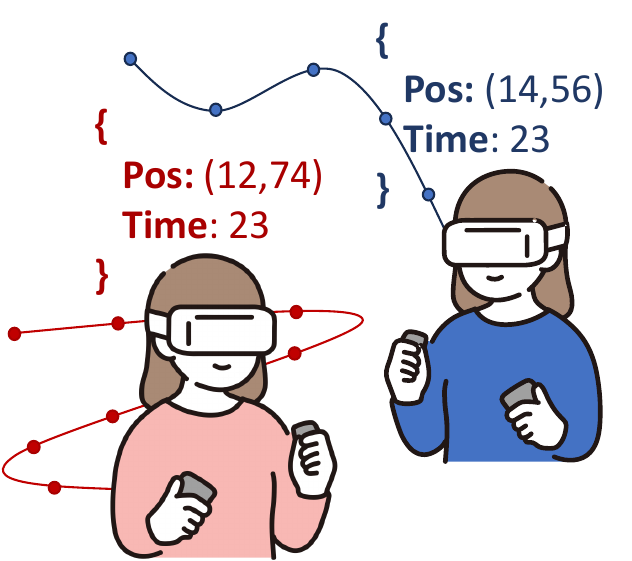}
        \Description{Two avatars whose trajectories are traced as dotted paths, each annotated with a record pairing a position with a timestamp, illustrating how co-presence and movements can be logged over time.}
        \caption{Social Logging.}
        \label{fig:social-logging}
    \end{subfigure}\hfill
    \begin{subfigure}[b]{0.19\textwidth}
        \centering
        \includegraphics[width=\linewidth,height=2.8cm,keepaspectratio]{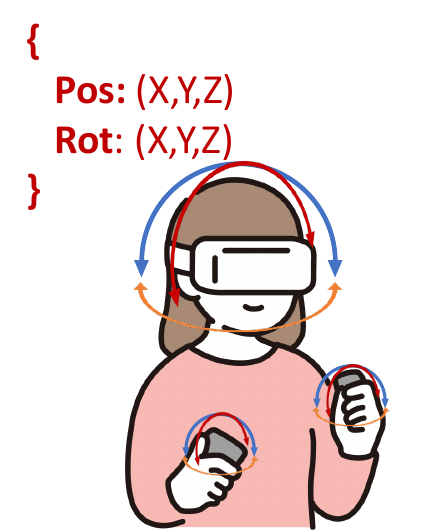}
        \Description{An avatar annotated with a record of position and rotation coordinates; arrows around the headset and both hand controllers mark the rotation axes tracked for motion-based fingerprinting.}
        \caption{FP with Motion Data.}
        \label{fig:movement-tracking}
    \end{subfigure}\hfill
    \begin{subfigure}[b]{0.19\textwidth}
        \centering
        \includegraphics[width=\linewidth,height=2.8cm,keepaspectratio]{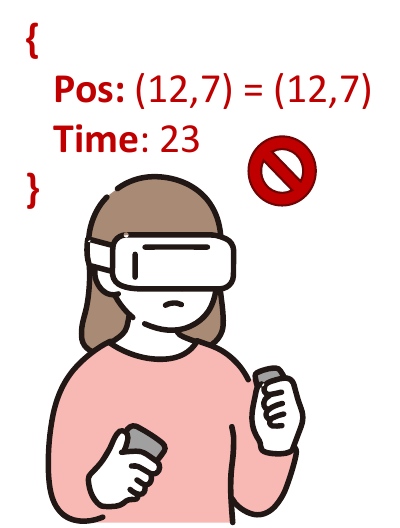}
        \Description{A frowning avatar next to a prohibition sign, annotated with a record whose position is repeatedly forced back to the same coordinates, leaving the user unable to move.}
        \caption{User DoS.}
        \label{fig:dos-attack}
    \end{subfigure}\hfill
    \begin{subfigure}[b]{0.19\textwidth}
        \centering
        \includegraphics[width=\linewidth,height=2.8cm,keepaspectratio]{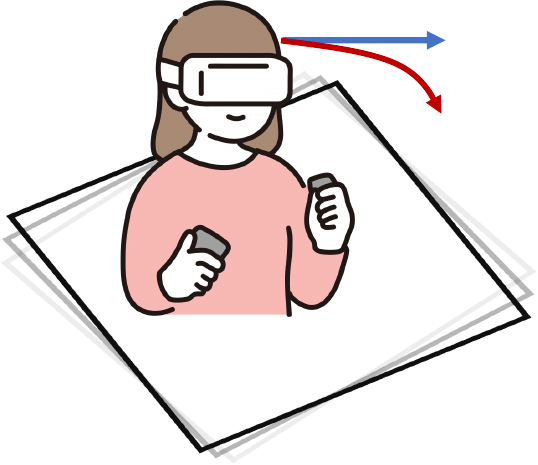}
        \Description{An avatar standing on a floor tile that the world gradually shifts and rotates; two arrows contrast the direction the user intends to move with the deviated path imposed on them.}
        \caption{Human Joystick.}
        \label{fig:human-joystick}
    \end{subfigure}
    \caption{Illustrative scenarios for state-of-the-art metaverse attacks.}
    \label{fig:attacks-overview}
\end{figure*}

Next, we analyze prior attacks targeting metaverse and VR platforms. Unlike the omniscience attacks of \Cref{sec:new_attacks}, these are reproductions of prior-literature attacks.
We show
that previously-proposed attacks, originally requiring either
(i)~a considerable level of technical expertise~\cite{Mengascini_Aurelio_Pellegrino_2024},
(ii)~the presence of specific software vulnerabilities~\cite{Vondrek_Baggili_Casey_Mekni_2023},
or (iii)~developer-level privileges~\cite{Casey_Baggili_Yarramreddy_2021}, can be
straightforwardly realized using only platform-provided world creation tools.
These attacks' technical requirements align with the
world-creation capabilities discussed in \Cref{subsec:capabilities}, which we
use for the attacks' re-implementation.

To ground our analysis in real-world threats, we surveyed \empirical{60}
academic papers on metaverse and VR security, identifying \empirical{32} that
propose attacks. We grouped these into seven
categories and focused on four:
sensitive data collection, denial of service, fingerprinting, and physical
manipulation. Attack categories relying on user deception, such as social
engineering or clickjacking, or on system-level access, such as keylogging,
were excluded from our analysis. 
\Cref{app:lit_attacks} 
summarize the survey, including categories, representative works, and inclusion criteria.
Below, we briefly
present the selected attack categories and corresponding attacks we
implemented.

\textbf{Sensitive data collection.} Both developers and users
can retrieve sensitive data about other users from virtual environments. For
example, developers can use their privileged access to collect and analyze user
positions, and hand and head
movements~\cite{Nair_Munilla_Garrido_Song_OBrien_2023}, e.g.,~to infer a user's
physical
space~\cite{Vilk_Molnar_Livshits_Ofek_Rossbach_Moshchuk_Wang_Gal_2015}. Users
can collect data by tampering with the client-side execution environment~\cite{Mengascini_Aurelio_Pellegrino_2024}. Since this is
the most prevalent category in our survey, we implement two
representative attacks: user re-identification and social logging.

\textbf{Denial of Service (DoS).} DoS attacks aim to disrupt the users' experience
or virtual environment functionalities. For example, both
developers~\cite{Cheng_Bhattacharya_Lin_Lee_Kumar_Tian_Kohno_Roesner} and
users~\cite{Mengascini_Aurelio_Pellegrino_2024} can encapsulate target objects
with other objects, making them inaccessible. %

\textbf{Fingerprinting.} Developers can use movement patterns
or device
characteristics~\cite{Munilla_Garrido_Nair_Song_2024,Nair_Rack_Guo_Wang_Li_Huang_Cull_OBrien_Latoschik_Rosenberg_et}
to track users across sessions, by creating persistent user profiles without
the users' knowledge or consent. Accordingly, we will demonstrate how we
can read and exfiltrate a user's motion data.

\textbf{Physical world.} These attacks exploit the virtual environment to
disrupt the user's physical experience or manipulate their actions. For
example, attackers can manipulate object appearances to redirect the user's
gaze~\cite{Ramirez_Spivack_DavidJohn_2024}, mismatch virtual and physical
movements to alter walking paths~\cite{Casey_Baggili_Yarramreddy_2021}, or
induce frame drops and erratic motion to cause
cybersickness~\cite{Valluripally_Gulhane_Hoque_Calyam_2022}. We implement
the human joystick attack~\cite{Casey_Baggili_Yarramreddy_2021} from this category.

\begin{figure}[th]
    \centering
    \includegraphics[width=0.7\columnwidth]{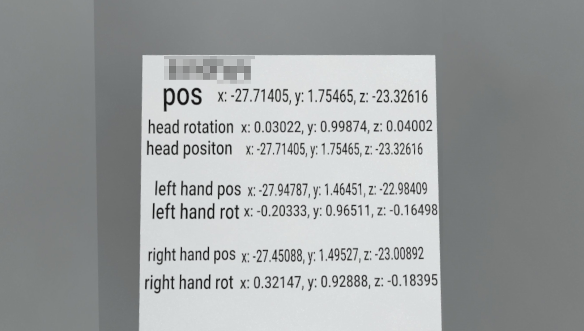}
    \Description{Screenshot of an in-world terminal displaying live tracking data of a target user as numeric coordinates: overall position, head position and rotation, and the positions and rotations of both hands. The username at the top is blurred.}
    \captionof{figure}{Control room terminal in Horizon Worlds, Personal Identifier blurred for anonymization.}
    \label{fig:control_room_attack}
\end{figure}

\subsection{Proof-of-Concept Attacks}
\label{sec:old_attacks}
 
We implement five representative state-of-the-art attacks
(\Cref{fig:attacks-overview}) across the selected platforms, to demonstrate how
world creators can trivially implement attacks from all major categories,
without requiring sophisticated technical skills or specific vulnerabilities,
as was the case in the past.
On Roblox, we reproduced several of these attacks with great ease using the platform's built-in AI coding assistant, which generated and attached the required scripts from a short prompt in a few attempts, without questioning their intent.

\textbf{User re-identification attack.} This attack links users across sessions
or worlds using persistent identifiers, and requires two capabilities: access
to user-identifiable information and a channel for data exfiltration. Frame's
no-code editor allows creators to embed user-specific information (e.g.,~emails
or usernames) into predefined request templates. 
This should trigger user
consent prompts (detailed in \Cref{app:in-app-editor}), but we found a
method to access identifiers directly via mesh objects without any
warnings.
Data can then be exfiltrated through client-side scripts using the
\lstinline{Fetch} function. Roblox exposes immutable user IDs and mutable
usernames; creators can exfiltrate them using built-in HTTP requests. Horizon
Worlds blocks network access, so we implement a hidden control room (see \Cref{fig:control_room_attack}) that
displays usernames on a canvas; the creator can join the world and use
Oculus casting with Selenium and OCR to extract data from rendered frames.
VRChat also restricts outbound requests to static allowlisted URLs, so we
log data locally on the client when a known username (the attacker) is
detected and extract it from our disk. In Spatial, identifiers are accessible
via the \lstinline{actorService} API; while HTTP access requires a
\$100/month subscription, logs can be written to the browser console and
scraped using Selenium, bypassing the paywall.

\textbf{Fingerprinting through motion data.} This attack builds motion profiles
from users' head and hand movements and requires access to spatial pose data,
identifiable user information, network access, and continuous execution. 
All platforms expose positional data for user avatars. For example, in Frame,
while HMD and controller positions are poorly documented, we used
\lstinline{GetAllMeshes} to identify the relevant meshes and extract position
and rotation.
Other platforms provide explicit APIs, such as \lstinline{leftHand.position} in
Roblox or \lstinline{GetTrackingData(Head)} in VRChat. For user identifiers and
exfiltration, we reused the techniques described in the re-identification attack. To
ensure continuous data collection, we relied on per-frame callbacks or looping
constructs;~e.g., \lstinline{while(true)} in Roblox,
\lstinline{OnBeforeRenderLoop()} in Frame,
and the visual \lstinline{while} block in Horizon Worlds.

\textbf{Social logging.} This attack involves monitoring users'
presence and spatial movements within a virtual world to infer social
behaviors. It requires continuous spatial data access, network communication
capabilities, and continuous execution of the logging script.
To access all players, we iterate over a parent object containing user
entities. Roblox, VRChat, and Spatial provide dedicated methods:
\lstinline{GetPlayers()}~, \lstinline{VRCPlayerApi.GetPlayers()}~, and
\lstinline{actorService.actors}~, respectively. In Horizon Worlds, we built a
user list by handling user join and leave events. In Frame, client-side scripts
start logging automatically upon world load, so no iteration is needed.

\textbf{Human joystick.}
This attack manipulates a user's position or view by gradually adjusting their
camera or movement direction, causing them to unknowingly follow a physical
path. The goal is to disorient users and influence real-world movement without
their awareness. It requires write access to camera or positional data, and
runs continuously to apply subtle, incremental changes over time.
In Frame, Spatial, and Roblox, we directly modified the camera rotation via
\lstinline{GetMeshByName(..).cameras[0].rotation},
\lstinline{avatar.rotation}, and \lstinline{CurrentCamera.CFrame},
respectively.
In VRChat, we used \lstinline{TeleportTo(pos, rot)} to control rotation without
altering position.
Horizon Worlds blocks direct pose manipulation unless “Enable Player Movement” (see \Cref{fig:meta_warning}) is activated, which should warn users, though we could not trigger the warning. To bypass this, we scripted a collidable platform tracking the user's feet and moving accordingly, affecting position but not rotation. This adaptation
allowed for human joystick behavior but is limited in scalability and precise directional control. For continuous execution, we used the same techniques as the fingerprinting attack.

\textbf{User DoS.} This attack disrupts a user's experience by immobilizing
them or repeatedly teleporting them to inaccessible locations, requiring
identification of the target, control over their movement, and continuous
execution. 
In VRChat, we use the built-in \lstinline{Immobilize()} method. In Roblox and
Spatial, we repeatedly teleport the user to a fixed vector using
\lstinline{humanoidRootPart.Position} and \lstinline{avatar.position}.
In Frame, we modify the parent node of the active camera's
position, effectively teleporting the user. Additionally, we freeze the client
with an infinite loop (\lstinline{while(true)}), halting all input.
Horizon Worlds does not allow direct pose changes unless “Enable Player
Movement” is enabled (with an untriggered warning); as an 
adaptation, we use
collidable objects to manipulate the player's position, catapulting them into the
sky or out of bounds. User targeting and execution persistence follow the
same approaches as prior attacks.

\section{Platform Protections}
\label{sec:countermeasures}

Having shown that world creators can achieve omniscient visibility and
control within their own worlds using only standard creation tools, we now
examine whether metaverse platforms provide effective safeguards against such
threats, or meaningful signals to affected users. This section evaluates the
protections platforms claim to offer, spanning vetting processes, in-world
safety mechanisms, runtime permissions and warnings, and creator-facing
policies, and weigh these claims with our empirical observations. %

\textbf{Vetting and review processes.} Several platforms describe some form of review prior to, or following, the
publication of worlds. To understand the scope of such processes, we reviewed
the public documentation of \empirical{14} platforms (see
\Cref{tab:metaverse_capabilities_with_discarded}), focusing on whether, and how,
world content is evaluated before being made accessible to other users.
Where review mechanisms are described, they appear primarily oriented toward
moderating visible or thematic content. For example, VRChat's Community Labs
review addresses inappropriate or offensive material~\cite{Iwanttom91:online};
Roblox requires creators to complete a ``Maturity \& Compliance Questionnaire''
centered on violence, fear, or crude humor~\cite{Howyouca33:online}; and
Horizon Worlds applies content ratings and policy checks aimed at adult or
harmful content~\cite{MetaforD94:online}. Community discussions similarly frame
these systems as content moderation tools rather than as mechanisms for
assessing world logic or technical
behavior~\cite{RobloxEx10:online,FalseDMC0:online}.

Across the entire duration of the experiments described in this paper, we did
not observe \emph{any} intervention, review feedback, or additional scrutiny
related to the technical behaviors implemented in our worlds. From our
perspective as creators and participants, publishing and interacting with these
worlds did not surface any signals suggesting that creator logic was being
examined beyond content-level considerations. While this does not rule out the
existence of internal checks, it suggests that any such processes are not
externally visible and, at minimum, do not manifest in ways that would
constrain or deter the behaviors of our attacks.

\textbf{In-world social safety mechanisms.} 
In addition to vetting, platforms emphasize in-world safety features intended
to give users control over their immediate social experience. Platforms such as
VRChat~\cite{VRChatSa61:online} and Horizon Worlds~\cite{Safetyan94:online}
allow users to mute or block others, hide avatars, or reduce visual and audio
effects, which are primarily designed to mitigate harassment or unwanted
user-to-user interactions. These controls operate at the level of interpersonal
interaction and do not constrain the behavior of world creators or the logic
embedded in a world. Consequently, they do not affect the attacks
presented in this paper, which are implemented through creator-controlled world
logic rather than direct user behavior.

\textbf{Permissions and user awareness.} 
A more direct approach to constraining world-creator behavior would involve
runtime permission systems or user-facing warnings shown when sensitive
capabilities are exercised.
Among the platforms we evaluated, only \emph{two} explicitly reference such
mechanisms. FrameVR presents a consent dialog when a creator attempts to access
a user's email address via its no-code action system
(\Cref{fig:frame_consent}); however, we found that the same information is also
accessible via JavaScript, allowing creators to bypass this warning entirely.
Horizon Worlds requires creators to enable a world-level setting to manipulate
player movement, and the editor indicates that enabling this option will notify
users (\Cref{fig:meta_warning}); however, we did not observe any corresponding
user-facing warning in practice. For the remaining platforms, we found no
indication of permission prompts or warnings related to creator access to
sensitive capabilities.

In our experiments, as creators we 
accessed and combined
sensitive inputs, such as spatial pose data, audio routing, and persistent user
identifiers, without 
triggering permission prompts or warnings for
affected users. In cases where platforms claim that opt-in settings or
notifications should be displayed, we did not 
observe them in practice. Consequently, users 
in our attack scenarios received
no indication that their behavior was being monitored, logged, or manipulated
beyond what is shown in a typical world experience.
Our observations are limited to externally visible behavior
and do not exclude the possibility of undocumented mechanisms. Nonetheless,
from a user's perspective, these systems did not provide actionable
transparency or awareness during our evaluation.

\textbf{Creator policies and terms.} 
Finally, metaverse platforms articulate expectations for creator behavior
through policies, terms of service, and SDK or analytics guidelines. To assess
how these documents address the behaviors enabled by our attacks, we reviewed
publicly available creator-facing policies for the five analyzed platforms and mapped explicit restrictions to
the techniques and outcomes of our attacks.

In general, policies are most explicit in restricting the collection of
sensitive or real-identity information, profiling or cross-service tracking,
and the export of user-derived data off-platform. Platforms such as Meta,
Roblox, and VRChat prohibit in-world collection of certain identifying
data~\cite{MetaTerms,MetaCreatorCoC,RobloxThirdPartyTerms,RobloxCommunityStandards,VRChatTOS}.
Profiling is prohibited on Roblox and conditionally permitted on other
platforms, typically contingent on notice or
consent~\cite{RobloxThirdPartyPolicy,SpatialPrivacyPolicy,VRChatTOS}. Exporting
user data is similarly constrained in VRChat's SDK License, Roblox's
third-party data policies, and Frame's
terms~\cite{VRChatSDKLicense,RobloxThirdPartyPolicy,FrameTerms}.

At the same time, these policies coexist with creation tools that continue to
expose the technical primitives required to implement the restricted behaviors.
Enforcement appears to rely primarily on terms-of-service obligations and
retrospective action, rather than on preventative technical controls. As a
result, responsibility for compliance is 
delegated to creators, while
users have limited visibility into how worlds
handle their data
or shape their experience.

\section{Discussion}
\label{sec:discussion}

Our work reveals systemic security and privacy risks introduced by the
world-creator model adopted by metaverse platforms. Here,
we distill cross-cutting lessons, discuss potential mitigations and
their challenges, and outline the scope and limitations of our study.

\subsection{Lessons Learned}

\hspace*{\parindent}\textbf{World creation fundamentally changes the trust model.} Metaverse platforms blur the traditional boundary between users and
developers by delegating powerful, developer-like capabilities to ordinary
users acting as world creators. While creators remain untrusted from the
platform's perspective, they are granted fine-grained control over runtime
logic, sensory cues, and interaction rules. This hybrid role does not fit
existing threat models,
introducing new
attacks that are neither
purely developer- nor user-driven. Crucially, crafting this attack logic demands far less skill than full application development, widening the pool of capable adversaries.

\textbf{World creators violate privacy and trust expectations.} 
Visual occlusion, spatial audio, and architectural metaphors implicitly signal privacy and exclusivity~\cite{SB2025WhatAT,williamson2021proxemics}, and even when privileged actors
override
these boundaries, users expect such access to be visible or signalled~\cite{bailey2025students}. This
matches popular systems such as Zoom and Slack, where administrators
configure spaces but do not silently monitor private exchanges. 
Our attacks
show that creator-built worlds can violate this trust model: ordinary creators
can silently observe, listen to, track, and manipulate users through standard
creation tools. 
The core risk is thus not creator control itself, but
unobservable creator authority that violates expectations of visible or
signalled access.

\textbf{Omniscience emerges from composition, not vulnerabilities.} Individually, most world-creation features appear benign. Collectively,
through creative design and synthesis, they
enable omniscient surveillance and manipulation. Across platforms and
tools, creators can recombine persistent execution, spatial state
access, and object control %
to achieve powerful attacks.
Importantly, these capabilities emerge without exploiting software vulnerabilities
or undocumented APIs.

\textbf{The barrier to impactful attacks is low.} Compared to prior metaverse attacks that required client tampering or network
analysis, the attacks in this paper can be implemented using visual editors or
simple scripts. Even when platforms restrict specific APIs, creators can often
reconstruct equivalent behaviors through alternative means. 
Additionally, this barrier is also dropping, as tooling evolves from block-based editors to AI assistants that synthesize attack logic from a prompt. This shifts the
threat model from expert attackers to, essentially, any motivated user. 

\textbf{Existing defenses are structurally misaligned.} Platform defenses focus on content moderation, user-to-user safety controls, or
contractual restrictions. These mechanisms do not meaningfully constrain
creator-controlled world logic. As a result, malicious behavior can persist
without triggering vetting, runtime warnings, or user awareness, even when it
violates stated policies.

\subsection{Mitigation Strategies and Open Challenges}

Our findings suggest that there is no single, standalone 
solution to the risks
introduced by creator-controlled worlds. Instead, mitigating these threats
requires a combination of complementary measures, each with its unique
challenges in the metaverse setting.

\textbf{Raising the entry barrier for creators.} One immediate mitigation is
to increase the cost of becoming a creator~\cite{RobloxPublishing:online}.  Other ecosystems, most notably
mobile app platforms, rely on account vetting, including government ID
verification~\cite{Identity55:online,GooglePl97:online}.
Similar mechanisms
could limit who can publish public worlds, or what capabilities are unlocked at
early stages. However, this directly conflicts with the low-friction,
creator-driven
 growth
model that metaverse platforms
 currently
favor.

\textbf{World vetting beyond visible content.} In principle, platforms could vet worlds in a manner analogous to mobile app
review, inspecting logic and behavior before publication. In practice, this
is substantially harder. Our attacks rely 
solely on standard features and
can be embedded inconspicuously in otherwise benign experiences. Unlike
mobile apps, the harmfulness of a world often depends on how logic unfolds
dynamically in a 3D, multi-user environment.

Traditional program analysis techniques struggle in this domain. Static
analysis cannot capture how spatial, visual, and auditory behaviors interact at
runtime. Dynamic analysis would require exploring a vast state space shaped by
user movement, interaction, and timing.
Even detecting components such as cameras or logging logic is insufficient, as these are common in legitimate worlds.

\textbf{Permissions and transparency help, but only to a point.} Runtime permissions or transparency labels (e.g., indicating audio routing,
data collection, or cameras) could improve user awareness.
However, many benign features would trigger frequent prompts, risking
habituation~\cite{anderson2016warning,238307}. Moreover, disclosures often remain under creator control, limiting
their effectiveness.
Designing meaningful, abuse-resistant permission systems remains an open problem.

\textbf{Reconsidering platform-enforced invariants.} A more fundamental approach may be to reassert certain platform-level
invariants, such as limits on sensory access or observation, that creators cannot
override. This would represent a shift away from fully malleable worlds toward
stronger guarantees of user protection, but would require careful trade-offs
with expressiveness and creativity.

\subsection{Limitations and Scope}
\label{subsec:limitations}

\hspace*{\parindent}\textbf{Platform selection.} Our study focuses on free-to-access metaverse platforms, 
where large-scale user activity occurs and aligning with prior security research.
Paid or enterprise-oriented metaverses may exhibit different
trade-offs, which we leave to future work. 
We also do not measure the prevalence of malicious worlds in the wild.
Understanding how often these attacks occur in
practice is an important direction
for future work.

\textbf{Platform updates.}
Metaverse platforms are rapidly evolving, and creator tooling changed 
after our study. For example, Meta introduced a desktop editor that
supports TypeScript, which in practice lowered the effort required to implement
several attacks; importantly, attacks developed using the in-app editor remain
feasible under the new tooling. Conversely, Frame deprecated its custom editor
and removed general scripting support after our experiments, rendering some
script-based attacks from \Cref{sec:old_attacks} infeasible. %
Nonetheless, while specific attacks may become easier or harder as tooling
changes, our core finding \emph{that powerful creator capabilities persist across
platforms and editor models} remains unchanged.

\section{Related Work}

This paper presents the first study on the risks of the world-creation
capabilities in metaverse platforms. 
Here, we discuss prior work on risks of user-generated content in different contexts, as well as prior security analyses of the metaverse through the lens of the well-established roles of developers and users.

\textbf{User-generated content.} A related line of work explores the
implications of allowing user-generated content (UGC) to be uploaded or
utilized in applications. UGC and creators are often seen as user-empowered
agents capable of producing harmful content or environments, particularly in
games. Zhang et al.~\cite{10.1145/3677076} and Kou et
al.~\cite{10.1145/3563657.3595960} have explored harmful patterns in UGC, such
as embedding problematic incentives like pervasive microtransactions in Roblox.
However, these studies focus on harmful design patterns rather than specific
attacks executable through the game, which has been widely covered by the
press, including themes like sexual content~\cite{RobloxTh2:online} and
terrorism~\cite{Robloxis71:online}. UGC can also serve as an attack vector,
enabling malicious actors to actually exploit systems. For instance, file
upload validations can be bypassed to upload unrestricted
content~\cite{Lee2020FUSEFF}, potentially leading to serious risks such as
remote code execution~\cite{Unrestri49:online}. While traditional attacks often
rely on encapsulating payloads in seemingly benign files, the metaverse
introduces a unique dimension where the payload is not just a file but the
actual logic of a virtual world. This distinction underscores the need to
study UGC-based threats in metaverse environments, where malicious
logic can use user experiences to launch attacks.

\textbf{Security threats in the metaverse.}
As part of our analysis of the risks introduced by the world creator role, we have also demonstrated how previously-proposed attacks can be trivially deployed. Next, we briefly discuss relevant prior work in the space.

\emph{Developers.} Developers create and maintain the platform applications,
typically leveraging game engines such as Unity~\cite{VRChatUnity:online} or
Unreal Engine~\cite{WhatEpic57:online}, or using custom-built
engines~\cite{webavers8:online}. They also manage the supporting
infrastructure, including servers, databases, and tools for world creation.
Developers possess significant control over the metaverse,
granting them control over both 
the application and its underlying infrastructure.
If malicious, developers can conduct various attacks. For example, they can
exploit immersive scenarios to redirect users' movements in the physical world~\cite{Casey_Baggili_Yarramreddy_2021}
or collect sensitive user data such as physical
characteristics~\cite{Munilla_Garrido_Nair_Song_2024} and movement
patterns~\cite{Nair_Guo_Mattern_Wang_OBrien_Rosenberg_Song_2023}. As operators
of the platform infrastructure, they can gather and analyze extensive user
data, potentially enabling surveillance~\cite{Trimananda_Le_Cui_Ho_Shuba_Markopoulou_2022}.
This threat has led researchers to investigate mitigations, such as
differential privacy mechanisms~\cite{Nair_MunillaGarrido_Song_2023}.

\emph{Users.} Users access the metaverse through client applications
that render virtual environments and synchronize with servers. Their attacks
typically exploit client-server vulnerabilities or overreach by the client
app. Common techniques include XSS to bypass security
mechanisms~\cite{Vondrek_Baggili_Casey_Mekni_2023}, client-side memory scanning
to capture sensitive information~\cite{Mengascini_Aurelio_Pellegrino_2024}, and
network tampering to infer personal
details~\cite{Su_Cai_Beeler_Dresel_Garcia_Grishchenko_Tian_Kruegel_Vigna}. When traditional vulnerabilities are absent, attackers resort to complex
methods such as observing hand gestures or gaze patterns to infer
information~\cite{Yang_Bhaskar_Sarwar_Zhao_Hwang_Zheng,
Wang_Zhan_Shan_Dai_Panoff_Wang_2024}. Recent defenses, such as anti-cheat
systems~\cite{TheVRCha32:online} and limiting sensitive information shared
among clients~\cite{Aboutthe26:online}, aim to reduce users' attack surface.

\section{Conclusion}

In this paper, we performed the first assessment of the security and privacy
risks introduced by metaverse world creators.  We showed that ordinary users
can create worlds that violate other users' spatial, visual, and auditory
privacy, and distort their environmental perception. This is achieved through
the creative use of official toolchains for fabricating malicious world
logic across platforms, without exploiting software vulnerabilities or
developer privileges. We demonstrated the significant threat posed by the
world creator role through concrete instantiations of novel attacks that enable
covert user surveillance and manipulation, and re-implementations of
previously-proposed attacks.  Our research further shows that existing defenses
are ineffective: vetting focuses on content rather than behavior, permission
systems and user-facing warnings are limited or absent, and users remain
unaware of invasive logic executing within worlds.  Overall, our findings
reveal a fundamental mismatch between users' privacy expectations and the
powers granted to world creators, highlighting an urgent need for stronger,
behavior-aware safeguards in metaverse platforms.

\begin{acks}
We would like to thank the anonymous reviewers and shepherd for their
valuable feedback. This work was supported by the National Science
Foundation under grants CNS-2211574 and CNS-2143363. Any opinions,
findings, conclusions, or recommendations expressed herein are those of
the authors, and do not necessarily reflect those of the NSF.
\end{acks}

\bibliographystyle{ACM-Reference-Format}
\balance
\bibliography{references}

\appendix %
\crefalias{section}{appendix}
\section{Open Science}
\label{sec:openscience_datasharing}
\hspace*{\parindent}\textbf{Data sharing.} We share in our repository the survey data for the platforms,
literature survey data on attacks, and the finalized codebooks with resolved
conflicts. 
The full
application-survey, the complete coding of the
systematic analysis, and the notes on how coding conflicts were resolved are all included in the artifact
repository.
All data, codebooks, resolution notes, and supporting materials are
available in the main artifact repository:
\url{https://github.com/andrea-mengascini/worldcreator-artifact/}.

\textbf{Attack artifacts.} Anonymized video demonstrations of all our attacks are available at
\url{https://andrea-mengascini.github.io/worldcreator-artifact/}. 
These videos include clear textual descriptions and timestamps for each phase of the
attacks. The same videos are also provided for offline download within the main
artifact repository.
The repository also includes the 3D model used to implement the one-way
material, enabling complete reproduction of this component of our evaluation.

\section{Ethical Considerations}

\label{sec:ethics}
We carefully considered ethical implications throughout the entire lifecycle of
this research, from the design of experiments to the potential impact of the
results. In this section, we outline our ethical considerations for each
stakeholder involved and the measures taken to mitigate potential risks.
Importantly, for ethical reasons, we did not upload any public
worlds containing attacks or malicious payloads to test the
platform vetting processes. Uploading such worlds could have exposed real users
to harm, and therefore we restricted all experimental evaluations to private or
unlisted instances under our exclusive control.

\textbf{Metaverse platform users.} We ensured that our experiments would not
inadvertently affect the users of the tested  metaverse platforms. To achieve this, all
tests were conducted in private or unlisted instances. These instances were restricted to our test accounts, ensuring public
users could not inadvertently join or be impacted. For platforms supporting
private environments, we utilized password-protected or invite-only instances.
On platforms that did not provide private instances, we created unlisted
instances and conducted all experiments exclusively with our test accounts.

After completing the experiments, we removed all test rooms except for those temporarily retained for reviewers, which can also be deleted and will be removed upon completion of the review cycle prior to final submission. In cases where deletion was not possible, we replaced private or unlisted instances with blank worlds stripped of all functionalities to eliminate any potential residual impact.

\textbf{Metaverse platform providers.} 
We adhered to the intended functionalities
by designing and uploading worlds using only official APIs and tools provided
by the platforms. Our experiments did not involve testing or exploiting
vulnerabilities in the platforms' tools or server infrastructure, ensuring
compliance with their operational integrity.

\textbf{Responsible disclosure.} 
When we identified possible attacks, we initiated a responsible disclosure process with the affected platforms, following the protocol detailed below. For each platform, we first requested a valid security contact point using available communication channels such as web forms, email, or official social media. After obtaining the appropriate contact, we shared a detailed technical report describing the issues and steps to reproduce the attacks.
Roblox replied through their bug bounty program, stating that
the reported issue does not constitute a security vulnerability according to
their criteria. This response further indicates that, in practice, malicious worlds
leveraging the techniques detailed in this work would pass the vetting process
of major platforms.
Meta similarly concluded that the reported behavior does not constitute a security or privacy issue, arguing that a world creator gains no additional information beyond what would be available through in-world proximity.
For the remaining platforms, we have notified them and did not receive a response.

\textbf{Artifact sharing.} To reduce the risk of misuse, we will not
publicly release executable attack scripts, exploit code, or ready-to-deploy
malicious world files while the underlying issues remain unmitigated. We only shared technical artifacts privately with reviewers, affected platform providers and, upon
request, we will share them with vetted researchers for research purposes.

\section{Platform Selection Methodology}
\label{app:platform-selection}

This appendix describes the methodology used to identify, filter, and rank metaverse platforms that support user-created worlds. The goal of this process was selecting a representative set of widely used, freely accessible platforms spanning different creation tool models and logic-coding paradigms.

\textbf{Candidates discovery}
We constructed an initial candidate list by surveying search engines.
First, we searched on Google using the keywords ``\texttt{VR social platform}'', ``\texttt{VR + Multiplayer}'', and ``\texttt{VR + Metaverse}'' and reviewed the pages of the first 100 results. Next, we performed keyword searches on two popular stores for virtual reality games: the Meta Quest Store~\cite{QuestStore:online} and the Steam Store~\cite{steam:online}. For the former, we used the search queries ``\texttt{Social}'' and ``\texttt{Metaverse}'', whereas for the latter we used ``\texttt{VR + Multiplayer}'' and ``\texttt{Metaverse}''. Then we processed up to 100 results from each store. 
Finally, we queried two unofficial databases of VR applications: VRDB~\cite{VRDBBrow49:online} and SteamDB~\cite{SteamDB34:online}, which provide data on Meta Quest and Steam games, respectively. For VRDB, we filtered for ``\texttt{Player-Multiplayer}'' applications, sorted by review count, and stored the first 50 free platforms classified as social multiplayer experiences. For SteamDB, we filtered for free platforms tagged as ``\texttt{Massively Multiplayer}'' or ``\texttt{Multiplayer}'' and ``\texttt{VR}'', sorted by online player count, and stored the first 50 entries. 
Across all sources, this process identified \empirical{561} candidate metaverses. After deduplication, \empirical{422} unique platforms remained.

\textbf{Candidates evaluation} We manually evaluated all \empirical{422} unique platforms by reviewing their store pages, official websites, and available documentation. Platforms were excluded if they did not satisfy all of the following criteria:
(1) native support for VR,
(2) multiplayer functionality,
(3) general availability to the public (i.e., not closed beta or discontinued),
(4) free to use.
After applying these criteria, \empirical{74} platforms remained.
Additionally, we excluded platforms that did not support user-generated content or lacked tools for creators. Among the \empirical{74}, \empirical{36} platforms provided explicit support for user-created worlds rather than limited object or scene customization.

\textbf{Ranking} To prioritize platforms with substantial real-world usage, we collected two popularity signals for each of the \empirical{36} platforms. First, we recorded the number of user reviews listed on the Meta Quest Store and/or Steam Store. When reviews were available on both stores, we retained the higher count. Second, we estimated platform website traffic using SimilarWeb\footnote{\url{https://www.similarweb.com/}}, which provides approximate monthly visitor counts for publicly accessible domains. While web traffic is an imperfect proxy for in-app usage, it offers a coarse indicator of overall platform adoption and visibility.
As these two metrics operate on different value ranges, we normalized them by calculating their combined standard scores $z$-comb-score = \((R_i - \mu(R))/\sigma(R) + (V_i - \mu(V))/\sigma(V)\),
where $\mu$ and $\sigma$ are the mean and standard deviation, and \(R_i\) and \(V_i\) represent the review count and visitor count for platform $i$, respectively. We then selected the first \empirical{25} platforms with the highest combined standard score.

\textbf{Metaverse selection and creation tools categorization}
\begin{table}[t]
    \centering
    \small	
    \begin{tabular}{l | ccc ccl | r}
    \toprule
     & \multicolumn{3}{c}{\textbf{Editor}} & \multicolumn{3}{c}{\textbf{Logic}} &  \\
    \cmidrule(lr){2-4} \cmidrule(lr){5-7}
    \textbf{Metaverse} & \rotatebox{90}{Game} & \rotatebox{90}{In-App} & \rotatebox{90}{Cust.} & \rotatebox{90}{No-code} & \rotatebox{90}{Code} & \rotatebox{90}{Lang} & \textbf{Rank}\\
    \midrule
        \rowcolor{gray!20} \textbf{Horizon Worlds}~\cite{MetaHori53:online} &  & \CIRCLE & \CIRCLE & \CIRCLE & \CIRCLE & TS & 3 \\

    \midrule
        \rowcolor{gray!20} \textbf{Frame}~\cite{Frame91:online}               &  & \CIRCLE & \LEFTcircle & \CIRCLE & \LEFTcircle & JS & 10 \\

    Resonite~\cite{Resonite25:online}            &  & \CIRCLE &  & \CIRCLE &  & - & 8 \\
    Neos VR~\cite{NeosMeta15:online}             &  & \CIRCLE &  & \CIRCLE &  & - & 13 \\
    \midrule
    \rowcolor{gray!20} \textbf{VRChat}~\cite{VRC:online}              & \CIRCLE &  &  & \CIRCLE & \CIRCLE & C\# & 4 \\
    \midrule
    Rec Room~\cite{RecRoom90:online}            & \CIRCLE & \CIRCLE &  & \CIRCLE & \CIRCLE & C\# & 6 \\
    \midrule
    \rowcolor{gray!20} \textbf{Spatial}~\cite{SpatialC52:online}             & \CIRCLE &  &  &  & \CIRCLE & C\# & 5 \\
    Viveport Verse~\cite{VIVERSEY41:online}      & \CIRCLE &  &  &  & \CIRCLE & JS & 7 \\
    Sansar~\cite{SansarOf99:online}              & \CIRCLE &  &  &  & \CIRCLE & C\# & 9 \\
    ChilloutVR~\cite{Chillout48:online}          & \CIRCLE &  &  &  & \CIRCLE & C\# & 12 \\
    Sinespace~\cite{Homesine38:online}           & \CIRCLE &  &  &  & \CIRCLE & Lua & 14 \\
    \midrule
    \rowcolor{gray!20} \textbf{Roblox}~\cite{HomeRobl54:online}              &  &  & \CIRCLE &  & \CIRCLE & Lua & 2 \\
    Overte~\cite{About—Ov22:online}              &  &  & \CIRCLE &  & \CIRCLE & JS & 18 \\
    Third Room~\cite{ThirdRoo77:online}          &  &  & \CIRCLE &  & \CIRCLE & C\# & 21 \\
    \midrule
    \multicolumn{8}{c}{\textit{Discarded platforms for the study:}} \\
    \midrule
    Multiverse~\cite{Multiver81:online}          &  & \CIRCLE &  & - & - & - & 15 \\
    Remio~\cite{RemioVR45:online}               &  & \CIRCLE &  & - & - &- & 17 \\
    GRAB~\cite{GRAB50:online}                     &  & \CIRCLE &  & - & - & - & 22 \\
    Penguin Paradise~\cite{SavaStud42:online}    &  & \CIRCLE &  & - & - & - & 23 \\
    DigiGods~\cite{DigiGods88:online}            &  & \CIRCLE &  & - & - & - & 24 \\
    \midrule
    Microsoft Mesh~\cite{Introduc87:online}      & - & - & - & - & - & - & 1 \\
    Somnium Space~\cite{SomniumS25:online}       & - & - & - & - & - & - & 11 \\
    Yeeps~\cite{YeepsHid1:online}                & - & - & - & - & - & - & 16 \\
    Villa~\cite{Metavers67:online}               & - & - & - & - & - & - & 19 \\
    AltspaceVR~\cite{Altspace36:online}          & - & - & - & - & - & - & 20 \\
    Derby~\cite{Throwbac23:online}               & - & - & - & - & - & - & 25 \\
    \bottomrule
    \end{tabular}
    \caption{Categorization of world creation tools. \LEFTcircle: Frame dropped their support for the Custom Editor which supported JavaScript after our experiments.}
    \label{tab:metaverse_capabilities_with_discarded}
\end{table}

Two researchers independently and iteratively
downloaded and reviewed tools, developer documentation, tutorials, and user
manuals to determine the high-level features of each platform's corresponding
world creation capabilities. 
We excluded \empirical{five} platforms (Multiverse, Remio, GRAB, Penguin Paradise, and DigiGods) because, while they provide visual editors for user-generated content, they do not support custom interaction logic via scripting or visual programming. We further excluded \empirical{six} platforms because they require paid subscriptions or additional fees for world creation access~\cite{Pricingf78:online} (Microsoft Mesh, Somnium Space, Yeeps, and Villa), do not support custom world creation (Derby), or have been permanently shut down~\cite{Altspace70:online} (AltspaceVR).
The remaining platforms form the final set analyzed in this paper. 
Overall, \empirical{seven} platforms extend existing game engine tools with
platform-specific SDKs. \empirical{Five} include built-in editors within their
metaverse applications, while another \empirical{five} offer external editors
as web or desktop apps. Two platforms (Horizon Worlds, Frame) support both in-app
and external world creation tools. The result of our analysis is shown in
\Cref{tab:metaverse_capabilities_with_discarded}.

\section{Literature Survey and Categorization}
\label{app:lit_attacks}
To contextualize our security analysis, we conducted a structured survey of prior work on attacks targeting VR and metaverse platforms.
We queried Google Scholar using combinations of the keywords ``\texttt{VR}'', ``\texttt{metaverse}'', ``\texttt{security}'', and ``\texttt{privacy}'', and iteratively followed references from relevant papers.
This process identified \empirical{60} papers published in high-impact venues, including \emph{IEEE S\&P}, \emph{USENIX Security}, \emph{NDSS}, and \emph{ACM CCS}.
Among these, \empirical{32} papers propose concrete attacks applicable to metaverse environments, conducted either by malicious developers or malicious users.

\setlength{\tabcolsep}{4pt} %

\begin{table}[t]
    \centering
    \small
         \begin{tabular}{l  c cccc c cc c}
         \toprule
         & \multicolumn{9}{c}{\textbf{Requirements}}\\
         \cmidrule(lr){2-10}
         & 
         \multirow{3}{*}[-1em]{\rotatebox{90}{{Network}}} & 
         \multicolumn{4}{c}{{SPD}} & 
         \multirow{3}{*}[-2.3em]{\rotatebox{90}{{UII}}} & 
         &
         &
         \multirow{3}{*}[0.3em]{\rotatebox{90}{{Cnt'd Exec.}}} \\

         \cmidrule(lr){3-6}
         & & \multicolumn{2}{c}{{Users}} & 
         \multicolumn{2}{c}{{Obj.}} 
         & 
         &
         \multicolumn{2}{c}{{Obj. Attr.}} &  \\
         
         \cmidrule(lr){3-4} \cmidrule(lr){5-6} \cmidrule(lr){8-9}
        \textbf{Attack Category} & 
        & \rotatebox{90}{R} & \rotatebox{90}{W} & 
        \rotatebox{90}{R} & \rotatebox{90}{W} & 
        & \rotatebox{90}{R} & \rotatebox{90}{W} & \\
        \midrule

        \multicolumn{2}{l}{\emph{Sensitive Data Collection}} \\
        \hspace{0.8em}Re-identification & \CIRCLE &  &  &  &  & \CIRCLE &  &  & \\
        \hspace{0.8em}Phys. world data coll. & \CIRCLE & \CIRCLE &  &  &  & \CIRCLE &  &  & \CIRCLE \\
        \hspace{0.8em}Pers. info. coll. & \CIRCLE & \CIRCLE &  &  &  & \CIRCLE &  &  & \CIRCLE \\
        \hspace{0.8em}Social logging & \CIRCLE & \CIRCLE &  &  &  & \CIRCLE &  &  & \CIRCLE \\
        \midrule

        \multicolumn{2}{l}{\emph{Denial of Service}} \\
        \hspace{0.8em}User DoS &  &  & \CIRCLE &  &  & \CIRCLE &  &  & \CIRCLE \\
        \hspace{0.8em}Object DoS &  &  &  &  & \CIRCLE & \CIRCLE &  & \CIRCLE & \CIRCLE \\
        \midrule

        \multicolumn{2}{l}{\emph{Fingerprinting}} \\
        \hspace{0.8em}Motion data & \CIRCLE & \CIRCLE &  &  &  & \CIRCLE &  &  & \CIRCLE \\
        \hspace{0.8em}User interaction & \CIRCLE & \CIRCLE &  &  &  & \CIRCLE &  &  & \CIRCLE \\
        \midrule

        \multicolumn{2}{l}{\emph{Physical}} \\
        \hspace{0.8em}Attention/gaze manip. &  & \CIRCLE &  &  & \CIRCLE &  & \CIRCLE &  & \CIRCLE \\
        \hspace{0.8em}Human Joystick &  & \CIRCLE & \CIRCLE &  &  & \CIRCLE &  &  & \CIRCLE \\
        \hspace{0.8em}Induced sickness &  &  & \CIRCLE &  &  & \CIRCLE &  & \CIRCLE & \CIRCLE \\
        \midrule
        \midrule        

        \emph{Keylogging} & \CIRCLE & \CIRCLE &  &  &  & \CIRCLE &  &  & \CIRCLE \\
        \midrule

        \emph{Clickjacking} &  &  &  & \CIRCLE & \CIRCLE &  &  & \CIRCLE & \\
        \midrule

        \multicolumn{2}{l}{\emph{Social Engineering}} \\
        \hspace{0.8em}Inception attack &  &  &  &  &  &  &  &  & \CIRCLE \\
        \hspace{0.8em}Impersonation &  &  &  &  &  & \CIRCLE &  &  & \\
        \bottomrule
    \end{tabular}
    \caption{Attacks in the literature grouped by category, with corresponding threat actor and required capabilities marked.}
    \label{tab:attack_groups_summary}
\end{table}

The identified attacks can be broadly grouped into \empirical{seven} attack categories, i.e., clickjacking attacks~\cite{Cheng_Bhattacharya_Lin_Lee_Kumar_Tian_Kohno_Roesner,Lee_Lee_Kim_Jana_Shin_Son_2021}, sensitive data collection attacks~\cite{Nair_Guo_Mattern_Wang_OBrien_Rosenberg_Song_2023,Jana_Molnar_Moshchuk_Dunn_Livshits_Wang_Ofek_2013,Roesner_Molnar_Moshchuk_Kohno_Wang_2014,Vilk_Molnar_Livshits_Ofek_Rossbach_Moshchuk_Wang_Gal_2015,Kim_Goutam_Rahmati_Kaufman_2023,Nair_Munilla_Garrido_Song_OBrien_2023,Guo_Dai_Luo_Zheng_Xu_He_2024,Zhang_Slocum_Chen_AbuGhazaleh_2023,Vondrek_Baggili_Casey_Mekni_2023,Mengascini_Aurelio_Pellegrino_2024,Shang_Chen_Wu_Yin_2022}, denial of service attacks~\cite{Cheng_Bhattacharya_Lin_Lee_Kumar_Tian_Kohno_Roesner,Mengascini_Aurelio_Pellegrino_2024,Cheng_Tian_Kohno_Roesner_2023}, fingerprinting attacks~\cite{Munilla_Garrido_Nair_Song_2024,Nair_Rack_Guo_Wang_Li_Huang_Cull_OBrien_Latoschik_Rosenberg_et,Tricomi_Nenna_Pajola_Conti_Gamberini_2023,Jarin_Duan_Trimananda_Cui_Elmalaki_Markopoulou_2024}, keylogging attacks~\cite{Wu_Shi_Zhang_Walker_Liu_Saxena_Chen_2023,Meteriz_2022,Ling_Li_Chen_Luo_Yu_Fu_2019,Slocum_Zhang_AbuGhazaleh_Chen_2023,Luo_Hu_Yan_2022,Wang_Zhan_Shan_Dai_Panoff_Wang_2024,Yang_Bhaskar_Sarwar_Zhao_Hwang_Zheng,Su_Cai_Beeler_Dresel_Garcia_Grishchenko_Tian_Kruegel_Vigna}, social engineering attacks~\cite{Yang_Li_Bhalla_Zhao,Vondrek_Baggili_Casey_Mekni_2023,Mengascini_Aurelio_Pellegrino_2024,Cheng_Bhattacharya_Lin_Lee_Kumar_Tian_Kohno_Roesner}, and physical attacks~\cite{Bonnail_Tseng_Mcgill_Lecolinet_Huron_Gugenheimer_2023,Ramirez_Spivack_DavidJohn_2024,Valluripally_Gulhane_Hoque_Calyam_2022,Casey_Baggili_Yarramreddy_2021,Tseng_Bonnail_McGill_Khamis_Lecolinet_Huron_Gugenheimer_2022}. 
For each attack, we analyzed the capabilities required for its realization in the context of our world-creator capability study based on platform-provided creation tools (\Cref{subsec:capabilities}), with the results summarized in \Cref{tab:attack_groups_summary}.

For the purposes of this paper, we excluded attack categories whose feasibility does not primarily depend on the capabilities of world creation tools.
For instance, social engineering attacks are excluded because their effectiveness relies on users being deceived, rather than on the technical capabilities of a creation tool. Similarly, keylogging attacks are excluded because modern systems restrict access to system keyboards while in use~\cite{Wang_Zhan_Shan_Dai_Panoff_Wang_2024} and sensor data, preventing side-channel exploitation. 
Finally, clickjacking attacks are also excluded as they require multi-application or multi-source setups where attackers exploit transparent overlays, interfaces from separate, malicious applications that invisibly capture user input, or embed malicious objects within benign ones, such as redirecting clicks to hidden ads.

\section{Victim-Side Attack Overhead}
\label{app:overhead}
To quantify the stealthiness claim of \Cref{sec:new_attacks}, we measured the
client-side overhead our attacks impose on the \emph{victim}. For each attack and
platform, we recorded the victim client's average frame rate, CPU utilization,
and GPU utilization with and without the attack running, and report the relative
changes in \Cref{tab:overhead}. We repeated each measurement three times, except for Spatial, which discontinued 3D World hosting for non-enterprise users. The impact is negligible: frame
rate and GPU utilization are essentially unchanged (on average $-0.13\%$ and
$+0.48\%$), while CPU utilization rises by only ${\sim}7.95\%$. The larger CPU
figures on Frame is a direct implication of the frame client-side execution model, where world logic
runs directly in the victim's browser.

\begin{table}[t]
\centering
\footnotesize
\setlength{\tabcolsep}{5pt}
\renewcommand{\arraystretch}{1.05}
\begin{tabular}{l r r r}
\toprule
\textbf{Attack} & $\Delta$FPS & $\Delta$CPU & $\Delta$GPU \\
\midrule
\multicolumn{4}{l}{\textit{VRChat}} \\
Parabolic Microphone   & $+0.18\%$  & $-2.23\%$  & $-0.24\%$ \\
Control Room           & $0.00\%$   & $+4.87\%$  & $+1.88\%$ \\
Astral Projection      & $+0.05\%$  & $+11.48\%$ & $+0.92\%$ \\
Unidirectional Material& $+0.16\%$  & $+1.58\%$  & $+2.45\%$ \\
Conversation Hijacking & $-0.05\%$  & $+1.69\%$  & $+2.96\%$ \\
\midrule
\multicolumn{4}{l}{\textit{Roblox}} \\
Parabolic Microphone   & $+1.59\%$  & $+6.82\%$  & $+1.44\%$ \\
Control Room           & $+0.44\%$  & $-6.79\%$  & $-2.94\%$ \\
Astral Projection      & $+0.92\%$  & $+4.73\%$  & $-2.18\%$ \\
Unidirectional Material& $-0.07\%$  & $+0.03\%$  & $-4.16\%$ \\
Conversation Hijacking & $+1.35\%$  & $+15.2\%$  & $+0.58\%$ \\
\midrule
\multicolumn{4}{l}{\textit{Spatial}} \\
Control Room           & $-0.07\%$  & $+10.8\%$  & $+2.20\%$ \\
Astral Projection      & $-0.14\%$  & $+2.89\%$  & $+0.98\%$ \\
Unidirectional Material& $-0.21\%$  & $-2.91\%$  & $+8.33\%$ \\
\midrule
\multicolumn{4}{l}{\textit{Frame}} \\
Astral Projection      & $-6.34\%$ & $+34.19\%$ & $-0.75\%$ \\
Unidirectional Material& $-0.05\%$ & $+39.82\%$ & $-3.13\%$ \\
\midrule
\multicolumn{4}{l}{\textit{Horizon Worlds}} \\
Astral Projection      & $-0.10\%$  & $+3.99\%$  & $-0.27\%$ \\
Unidirectional Material& $+0.07\%$  & $+4.67\%$  & $+1.25\%$ \\
\midrule
\textbf{Average}       & $-0.13\%$  & $+7.95\%$  & $+0.48\%$ \\
\bottomrule
\end{tabular}
\caption{Victim-side overhead of our attacks: relative change in average frame
rate, CPU, and GPU utilization on the victim client when the attack is active.}
\label{tab:overhead}
\end{table}

\section{Examples of In-App Editor Features}
\label{app:in-app-editor}
This appendix provides examples of creator-facing editor features that are intended to regulate access to sensitive capabilities or inform users about potential privacy implications.

\begin{figure}[H]
  \centering
  \includegraphics[width=0.85\columnwidth]{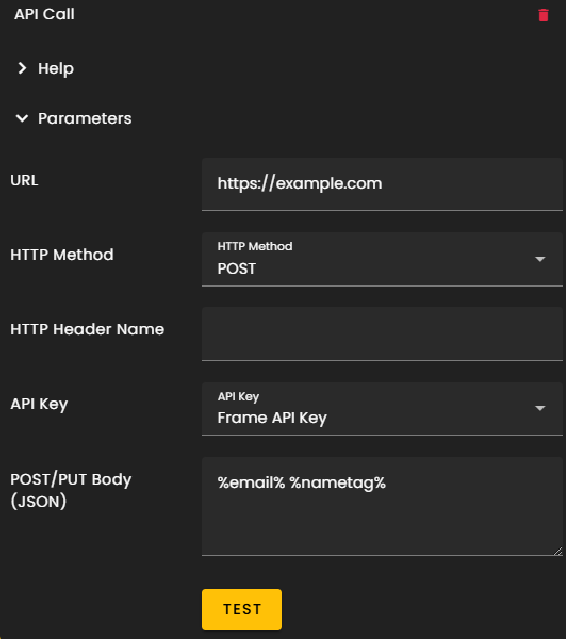}
  \Description{Screenshot of the API Call element in the Frame editor, with configurable fields for the target URL, HTTP method, header name, API key, and a POST body in which e-mail and nametag placeholders are inserted.}
  \caption{Frame in-app editor for no-code HTTP requests.}
  \label{fig:frame_network}
\end{figure}

\Cref{fig:frame_network} shows the dialog from Frame's in-app editor that allows creators to define external HTTP requests as part of no-code interaction logic. 
Through this interface, creators can transmit user associated data, such as email addresses and nametags, to external endpoints.

\begin{figure}[H]
  \centering
  \includegraphics[width=0.85\columnwidth]{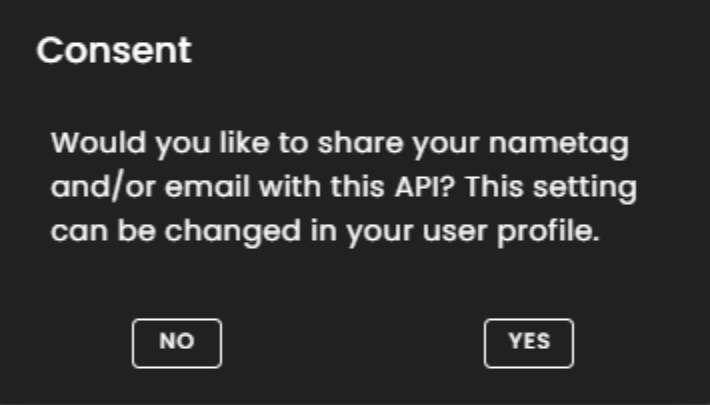}
  \Description{Screenshot of the consent dialog shown by Frame, asking whether the user would like to share their nametag and/or e-mail address with the API configured in the world, with No and Yes buttons.}
  \caption{Frame warning for user consent.}
  \label{fig:frame_consent}
\end{figure}

\Cref{fig:frame_consent} depicts the consent dialog shown to users when a Frame creator attempts to access the user's registered email address via the previous no-code action system. While this dialog suggests explicit user awareness and consent, we found that equivalent information could be accessed via JavaScript-based mechanisms without triggering this prompt.

\begin{figure}[H]
  \centering
  \includegraphics[width=0.85\columnwidth]{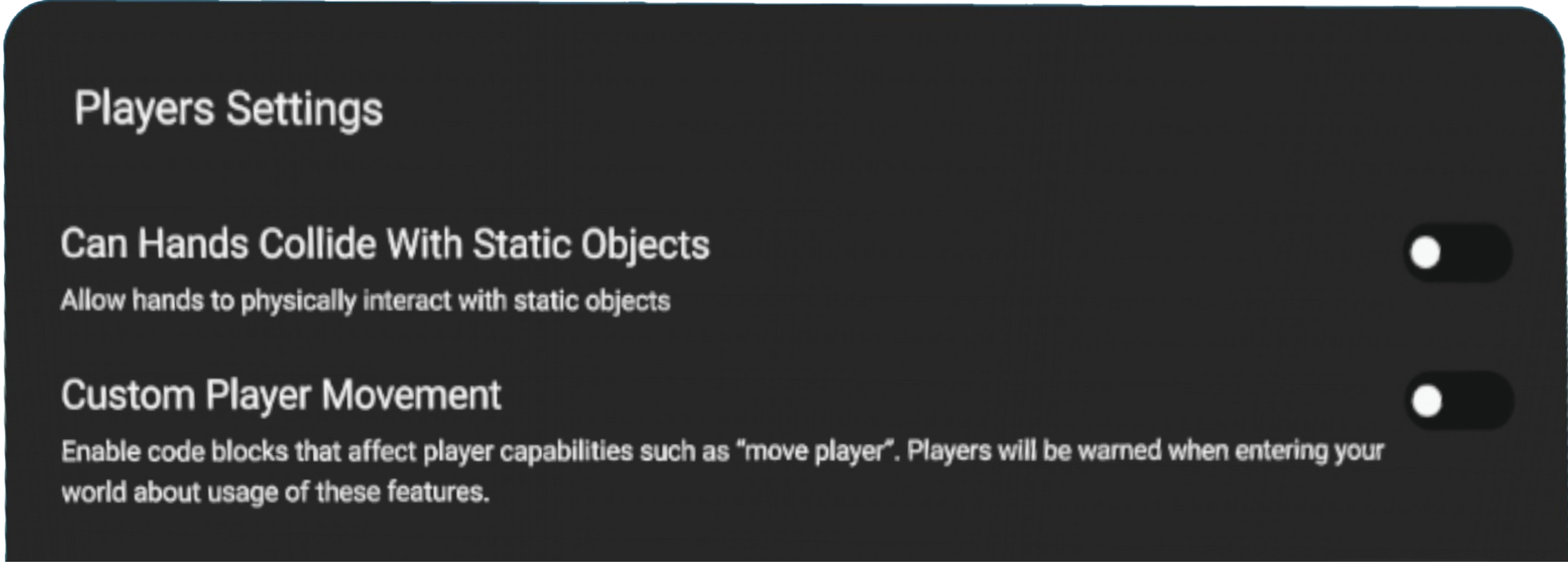}
  \Description{Screenshot of the Players Settings panel in the Horizon Worlds editor. The Custom Player Movement toggle enables code blocks that affect player capabilities, and its description states that players will be warned about the usage of these features when entering the world.}
  \caption{Horizon Worlds notice about user warning.}
  \label{fig:meta_warning}
\end{figure}

\Cref{fig:meta_warning} shows the creator-facing setting in Horizon Worlds required to enable custom player movement. The editor menu indicates that activating this option will notify users about potential movement manipulation. However, during our experiments, we did not observe any corresponding user-facing warning when this capability was selected.

\end{document}